\documentclass[12pt,preprint]{aastex}

\usepackage{lineno}

\slugcomment{}

\shorttitle{{\it Fermi}'s FSRQs}
\shortauthors{Ajello et al.}

\usepackage{calrsfs,euscript,mathrsfs}

\begin{document}


\title{The Luminosity Function of {\it Fermi}-detected Flat-Spectrum Radio
Quasars} 


\author{
M.~Ajello\altaffilmark{2,1}, 
M.~S.~Shaw\altaffilmark{2,1}, 
R.~W.~Romani\altaffilmark{2,1}, 
C.~D.~Dermer\altaffilmark{3},
L.~Costamante\altaffilmark{2}, 
O.~G.~King\altaffilmark{4}, 
W.~Max-Moerbeck\altaffilmark{4}, 
A.~Readhead\altaffilmark{4}, 
A.~Reimer\altaffilmark{5,2}, 
J.~L.~Richards\altaffilmark{4}, 
M.~Stevenson\altaffilmark{4}
}
\altaffiltext{1}{Corresponding authors: M.~Ajello, majello@slac.stanford.edu; R.~W.~Romani, rwr@astro.stanford.edu; M.~S.~Shaw, msshaw@stanford.edu.}
\altaffiltext{2}{W. W. Hansen Experimental Physics Laboratory, Kavli Institute for Particle Astrophysics and Cosmology, Department of Physics and SLAC National Accelerator Laboratory, Stanford University, Stanford, CA 94305, USA}
\altaffiltext{3}{Space Science Division, Naval Research Laboratory, Washington, DC 20375-5352, USA}
\altaffiltext{4}{Cahill Center for Astronomy and Astrophysics, California Institute of Technology, Pasadena, CA 91125, USA}
\altaffiltext{5}{Institut f\"ur Astro- und Teilchenphysik and Institut f\"ur Theoretische Physik, Leopold-Franzens-Universit\"at Innsbruck, A-6020 Innsbruck, Austria}

%
%
%
%
%
%

%
%
%
%
%
%
%
%
%
%
%

\begin{abstract}
{\it Fermi} has provided the largest sample of $\gamma$-ray selected
blazars to date. In this work we use a complete sample of FSRQs detected
during the first year of operation to  determine
the luminosity function (LF)  and its evolution with cosmic time.
The number density of FSRQs grows dramatically up to redshift $\sim$0.5--2.0 
and declines thereafter. The redshift of  the peak in the density is luminosity dependent,
with more luminous sources peaking at earlier times; thus the LF of $\gamma$-ray 
FSRQs follows a luminosity-dependent density evolution similarly to that of radio-quiet AGN. Also using data from the {\it Swift} Burst Alert Telescope
we derive the average spectral energy distribution of FSRQs in 
the 10\,keV--100\,GeV band and show that there is no 
correlation of the peak $\gamma$-ray luminosity with $\gamma$-ray peak frequency.
The coupling of the SED and LF allows us to predict that the 
contribution of FSRQs to the  {\it Fermi} isotropic $\gamma$-ray background is 
9.3$^{+1.6}_{-1.0}$\,\% ($\pm$3\,\% systematic uncertainty)
in the 0.1--100\,GeV band.
Finally we determine the LF of unbeamed FSRQs, finding that FSRQs have
an average Lorentz factor of  $\gamma=11.7^{+3.3}_{-2.2}$,
that most are seen
within 5$^{\circ}$ of the jet axis, and that they  represent only 
$\sim$0.1\,\% of the parent population.
\end{abstract}


\section{Introduction}

	The detection of luminous quasars at redshift $>$6 
\cite[e.g.][]{fan03,willott10} provides evidence of super-massive black hole
(SMBHs) formation in the first 1\,Gyr of cosmic time. There are
appreciable challenges to forming \cite[see e.g.][]{wyithe03,volonteri05b,begelman06,mayer10} 
and fueling \cite[see][]{kauffmann00,wyithe03,croton06} these objects at
such early times, although it is widely believed that strong accretion can
be initiated by major mergers \cite[see][]{kauffmann00,wyithe03,croton06}.

	Blazars represent an extreme manifestation of such AGN activity,
with radiation along the Earth line-of-sight dominated by a
relativistic jet. It is, as yet, unclear how such jet activity connects
with the more isotropically emitted bulk accretion luminosity. For
example according to \cite{blandford77}, the energy stored in a black
hole's spin can be extracted in the form of a relativistic jet. Thus blazar
evolution may be connected with the cosmic evolution of the spin states
of massive black holes. Radio-loud (jet dominated) blazars have been seen
at redshifts as high as z=5.5 \citep{romani06}, and it is plausible
that major mergers, more frequently experienced in the early universe,
might preferentially produce maximally rotating black holes \cite[e.g. see][]{escala04,dotti07}.

	Thus the study of radio-loud (RL) AGN, blazars, with strong 
relativistically beamed jets can provide a method to study jet activity,
BH spin, and major merger events. This can be done by determining the luminosity
function of blazars (LF, essentially the number of blazars per comoving
volume element within a certain luminosity range) and 
its evolution with redshift.  The {\it Fermi} Gamma-ray Space Telescope
provides one of the largest data sets with which to study the properties of blazars.
Thanks to its sensitivity and uniform coverage of the sky, {\it Fermi} has
detected hundreds of blazars from low redshifts out to  z=3.1 \citep{2LAC}.

The LF of blazars also allows us to evaluate their
contribution to the diffuse backgrounds and to
determine their relationship with
the parent population \citep{ajello08c,inoue11b}.
Blazars have been extensively studied at radio \citep{dunlop90,wall05},
soft X-ray \citep{giommi94,rector00,wolter01,caccianiga02,beckmann03,padovani07} 
and GeV energies  \citep{hartman99}.  It seems clear that flat spectrum
radio quasars (FSRQs) evolve positively \citep[i.e. there were more blazars 
in the past,][]{dunlop90} up to a redshift cut-off which depends on luminosity 
\citep[e.g.][]{padovani07,wall08b,ajello09b}.  In this respect FSRQs evolve
similarly to the population of  X--ray selected, radio-quiet, AGNs 
\citep{ueda03,hasinger05,lafranca05}.
On the other hand, the evolution of the other major class of 
{\it Fermi}-detected AGN,
BL Lac objects, and their relation to FSRQs, remains a matter of debate, with claims of no evolution
\citep{caccianiga02,padovani07} or even negative evolution \citep[e.g.][]{rector00,beckmann03}. Samples with larger redshift completeness fractions are needed
to study the LF of these claims.

In this work we report on the LF of FSRQs detected by {\it Fermi} in its first 
year of operation. There have been attempts in the past \citep[e.g.][]{chiang95}
to characterize
the evolution of $\gamma$-ray AGN, starting from the EGRET sample \citep{hartman99}.
One challenge was the small sample size and redshift incompleteness
of the EGRET set. Often it was assumed that the $\gamma$-ray detected blazars
had a LF following that of another band, e.g. radio- or X-ray selected blazars.
The results reported in e.g. \cite{stecker96}, \cite{narumoto06},
\cite{inoue09}, \cite{stecker11}
follow this approach.  Alternatively, a LF may be estimated from the
$\gamma$-ray sample directly \citep{chiang98,muecke00,dermer07,bhattacharya09},
although only a small ($\sim$60) blazar sample including both BL Lac objects
and FSRQs   was available (from EGRET data)
with  acceptable incompleteness.

We report here on a detailed LF measured from a sample of 186 $\gamma$-ray selected
FSRQs detected by {\it Fermi}. 
The work is organized as follows. Sections \ref{sec:sample} and \ref{sec:method} 
describe the properties of the sample used and the method employed to 
determine the LF of blazars. The luminosity function of FSRQs is derived 
in Section~\ref{sec:results}. In Section \ref{sec:sed} the spectral
energy distributions (SEDs) of {\it Fermi}'s FSRQs are analyzed in detail,
testing for possible correlations of the peak energy with the peak luminosity.
The contribution of FSRQs to the isotropic gamma-ray background\footnote{The 
isotropic gamma-ray background refers to the isotropic component
of the {\it Fermi} sky \citep{lat_edb} and as such might include 
components generated locally \citep[e.g.][]{keshet04} and components
of truly extragalactic origin.}\cite[IGRB, see][]{lat_edb} is determined and discussed in Section~\ref{sec:egb}.
Throughout this paper, we assume a standard concordance cosmology 
(H$_0$=71\,km s$^{-1}$ Mpc$^{-1}$, $\Omega_M$=1-$\Omega_{\Lambda}$=0.27).

\section{The Sample}
\label{sec:sample}
The First Fermi LAT Catalog \citep[1FGL,][]{cat1} reports on
more than 1400 sources  detected by {\it Fermi}-LAT during 
its first year of operation. The first LAT AGN catalog \citep[1LAC,][]{agn_cat}
associates $\sim$700 of the high-latitude 1FGL sources ($|$b$|\geq10^{\circ}$)
with AGN of various types, most of which are blazars.
The sample used for this analysis consists of sources detected by the pipeline developed
by \cite{pop_pap} with a test statistic (TS) significance greater than 50 
and with $|$b$|\geq$15$^{\circ}$.  For these sample cuts we have produced a set
of Monte Carlo simulations that can be used to determine and account for
the selection effects. This sample contains 483 objects, 186 of which are classified as
FSRQs. The faintest identified FSRQ has a 100\,MeV -- 100\,GeV band flux
of F$_{100}\approx10^{-8}$ ph cm$^{-2}$ s$^{-1}$. To limit the incompleteness 
(i.e. the fraction of sources without an association) we apply this as a flux
limit to the sample of 483 objects, 
resulting in a full sample of 433 sources of which 29
(i.e. $\sim$7\,\%) do not have associated counterparts.

The composition of this sample  is reported in Table~\ref{tab:sample}.
The 186 FSRQs detected by {\it Fermi} with TS$\geq50$, $|$b$|\geq15^{\circ}$,
and F$_{100}\geq10^{-8}$ ph cm$^{-2}$ s$^{-1}$ constitute the sample
that will be used in this analysis.
\begin{deluxetable}{lc}
\tablewidth{0pt}
\tablecaption{Composition of the $|$b$|\geq$15$^{\circ}$, TS$\geq$50, 
F$_{100}\geq10^{-8}$ ph cm$^{-2}$ s$^{-1}$ sample used in this
analysis.
\label{tab:sample}}
\tablehead{
\colhead{CLASS} & \colhead{\# objects }}
\startdata
Total                         & 433 \\
FSRQs                         & 186\\
BL Lacs                       & 157\\
Pulsars                       & 28 \\
Other\tablenotemark{a}          & 16 \\ 
Radio Associations\tablenotemark{b}          & 17 \\ 
Unassociated  sources         & 29 \\  

\enddata
\tablenotetext{a}{Includes starburst galaxies, LINERS,
narrow line Seyfert 1 objects and Seyfert galaxy candidates.}
\tablenotetext{b}{{\it Fermi} sources with a radio counterpart, but no
optical type or redshift measurement.}
\end{deluxetable}

\section{Method}
\label{sec:method}
A classical approach to derive the LF is based on the 1/V$_{\rm MAX}$ method 
of \cite{schmidt68} applied to redshift bins.  However, this method is known to 
introduce bias if there is significant evolution within the bins. Moreover, given our relatively 
small sample size and the large volume and luminosity range spanned, binning would result 
in a loss of information.  Thus we decided to apply the maximum-likelihood (ML) algorithm 
first introduced by \cite{marshall83} and used recently by \cite{ajello09b} for the study
of blazars detected by {\it Swift}.  The aim of this analysis is to determine the space
density of FSRQs as a function of rest-frame 0.1--100\,GeV luminosity
(L$_{\gamma}$), redshift (z) and photon index ($\Gamma$), by fitting to the functional form
\begin{equation}
\label{eq:1}
\frac{d^3 N}{dL_{\gamma} dz d\Gamma}= \frac{d^2N}{dL_{\gamma} dV}\times 
\frac{dN}{d\Gamma} \times \frac{dV}{dz} 
= \Phi(L_{\gamma},z) \times 
\frac{dN}{d\Gamma} \times \frac{dV}{dz}
\end{equation}
where $\Phi(L_{\gamma},z)$ is the luminosity function,  and $dV/dz$ is the co-moving volume element per unit redshift and unit solid 
angle \citep[see e.g.][]{hogg99}. The function $dN/d{\Gamma}$ is the (intrinsic) 
photon index  distribution and is assumed to be independent of z. It is modeled as a Gaussian:
\begin{equation}
\frac{dN}{d\Gamma} = e^{-\frac{ (\Gamma-\mu)^2}{2\sigma^2}}
\end{equation}
where $\mu$ and $\sigma$ are respectively the mean and the dispersion
of the Gaussian distribution.

 The best-fit LF is found by comparing, through a maximum-likelihood estimator, the number of expected objects (for a given model LF) to the observed
number while accounting for selection effects in the survey. 
In this method, the space of luminosity, redshift, and photon index is divided 
into small intervals of size $dL_{\gamma}dzd\Gamma$. In each
element, the expected number of blazars with luminosity $L_{\gamma}$,
redshift $z$ and photon index $\Gamma$ is:
\begin{equation}
\lambda(L_{\gamma},z, \Gamma)dL_{\gamma}dz d\Gamma   = 
\Phi(L_{\gamma},z)\Omega(L_{\gamma},z,\Gamma)\frac{dN}{d\Gamma}  \frac{dV}{dz} dL_{\gamma} dz d\Gamma
\label{eq:lambda}
\end{equation}
where $\Omega(L_{\gamma},z,\Gamma)$ is the sky coverage and represents the probability of 
detecting in this survey a blazar with luminosity $L_{\gamma}$,
redshift $z$ and photon index $\Gamma$. This probability was derived for 
the sample used here by \cite{pop_pap} and the reader is referred to that
aforementioned paper for more details. With sufficiently fine sampling of
the $L_{\gamma}-z-\Gamma$ space the infinitesimal element will either contain 0 or 1 FSRQs.
In this regime one has a likelihood function based on joint Poisson
probabilities:
\begin{equation}
L = \prod_i \lambda(L_{\gamma,i},z_i,\Gamma_i) dL_{\gamma} dz  d\Gamma
e^{-\lambda(L_{\gamma,i},z_i,\Gamma_i) dL_{\gamma} dz d\Gamma}
\times \prod_j e^{-\lambda(L_{\gamma,j},z_j,\Gamma_j) dL_{\Gamma} dz d\Gamma}
\end{equation}
This is the combined probability of observing one blazar in each bin of $(L_{\gamma,i},z_i,\Gamma_i)$ populated by one {\it Fermi} FSRQ and 
zero FSRQs for all other $(L_{\gamma,j},z_j,\Gamma_j)$.
Transforming to the standard expression $S=-2\ln\ L$ and dropping
terms which are not model dependent, we obtain:
\begin{equation}
S = -2\sum_i  {\rm \ln} \frac{d^3 N}{dL_{\gamma} dz d\Gamma}+ 2 
\int^{\Gamma_{max}}_{\Gamma_{min}} \int^{L_{\gamma,max}}_{L_{\gamma,min}} 
\int^{z_{max}}_{z_{min}} \lambda(L_{\gamma},\Gamma,z) dL_{\gamma} dz d\Gamma
\label{eq:s}
\end{equation}
The limits of integration of Eq.~\ref{eq:s}, unless otherwise stated, are:
$L_{\gamma,min}$=$10^{44}$\,erg s$^{-1}$, $L_{\gamma,max}$=10$^{52}$\,erg s$^{-1}$, $z_{min}$=0, $z_{max}$=6, $\Gamma_{min}=$1.8 and $\Gamma_{max}=$3.0.
The best-fit parameters are determined by minimizing\footnotemark{}
\footnotetext{The MINUIT minimization package, embedded in ROOT (root.cern.ch),
has been used for this purpose.}
$S$ and the associated 1\,$\sigma$ error are computed by varying the parameter
of interest, while the others are allowed to float, until an increment
of $\Delta S$=1 is achieved. This gives an estimate of the 68\,\%
confidence region for the parameter of interest \citep{avni76}.
While computationally intensive, Eq.~\ref{eq:s} has the advantage that each 
source has its appropriate individual detection efficiency and k-correction
treated independently.

In order to test whether the best-fit LF provides a good description
of the data we compare the {\it observed} redshift, luminosity, index
and source count distributions against the prediction of the LF.
The first three distributions  can be obtained from the LF as:
\begin{eqnarray}
\frac{dN}{dz} & = & \int^{\Gamma_{max}}_{\Gamma_{min}} \int^{L_{\gamma,max}}_{L_{\gamma,min}} \lambda(L_{\gamma},\Gamma,z) dL_{\gamma}  d\Gamma \\
\frac{dN}{dL_{\gamma}} & = & \int^{\Gamma_{max}}_{\Gamma_{min}} 
\int^{z_{max}}_{z_{min}} \lambda(L_{\gamma},\Gamma,z) dz d\Gamma \\
\frac{dN}{d\Gamma}  & = &  \int^{L_{\gamma,max}}_{L_{\gamma,min}} 
\int^{z_{max}}_{z_{min}} \lambda(L_{\gamma},\Gamma,z) dL_{\gamma} dz 
\end{eqnarray}
where the extremes of integrations are the same as in Eq.~\ref{eq:s}.
The source count distribution can be derived as :
\begin{equation}
N(>S) =  \int^{\Gamma_{max}}_{\Gamma_{min}} 
\int^{z_{max}}_{z_{min}}  \int^{L_{\gamma,max}}_{L_{\gamma}(z,S)}
 \Phi(L_{\gamma,z})\frac{dN}{d\Gamma}  \frac{dV}{dz} d\Gamma dz dL_{\gamma}
\label{eq:logn}
\end{equation}
where $L_{\gamma}(z,S)$ is the luminosity of a source at redshift $z$
having a flux of $S$.

To display the LF we rely on the ``N$^{obs}$/N$^{mdl}$'' method
devised by \cite{lafranca97} and \cite{miyaji01} and 
employed in several recent works \citep[e.g.][]{lafranca05,hasinger05}.
Once a best-fit function for the LF has been found, it is possible
to determine the value of the observed LF in a given bin of luminosity
and redshift:
\begin{equation}
\Phi(L_{\gamma,i},z_i) = \Phi^{mdl}(L_{\gamma,i},z_i) \frac {N^{obs}_i}{N^{mdl}_i}
\end{equation}
where $L_{\gamma,i}$ and $z_i$ are the luminosity and redshift of the i$^{th}$
bin, $\Phi^{mdl}(L_{\gamma,i},z_i)$ is the best-fit LF model and $N^{obs}_i$ 
and $N^{mdl}_i$ are the observed and the predicted number of FSRQs in that bin. 
These two techniques (the \cite{marshall83} ML method and the ``N$^{obs}$/N$^{mdl}$'' 
estimator) provide a minimally biased estimate of the luminosity function,
\citep[cf.][]{miyaji01}.

\section{Results}
\label{sec:results}

\subsection{Pure Luminosity Evolution and the Evidence for a Redshift Peak}
\label{sec:ple}

The space density of radio-quiet AGNs is known to be maximal at intermediate
redshift. The epoch of this `redshift peak' correlates with source luminosity 
\citep[e.g.][]{ueda03,hasinger05}. This peak may represent the combined
effect of SMBH growth over cosmic time and a fall-off in fueling activity
as the rate of major mergers decreases at late times.  To test whether 
such behavior is also typical of the LAT FSRQ population, we perform a fit to the 
data using a pure luminosity evolution (PLE) model of the form:

\begin{equation}
\Phi(L_{\gamma},z) = \Phi(L_{\gamma}/e(z))
\label{eq:ple}
\end{equation}
where
\begin{equation}
\Phi(L_{\gamma}/e(z=0)) = \frac{dN}{dL_{\gamma}}=\frac{A}{\ln(10)L_{\gamma}}
\left[\left(\frac{L_{\gamma}}{L_{*}}\right)^{\gamma_1}+
\left(\frac{L_{\gamma}}{L_{*}}\right)^{\gamma2} 
\right]^{-1}
\label{eq:ple1}
\end{equation}
and 
\begin{equation}
e(z) = (1+z)^k e^{z/\xi}.
\label{eq:ple2}
\end{equation}

\begin{figure*}[ht!]
  \begin{center}
  \begin{tabular}{ccc}
\hspace{-1cm}
    \includegraphics[scale=0.32]{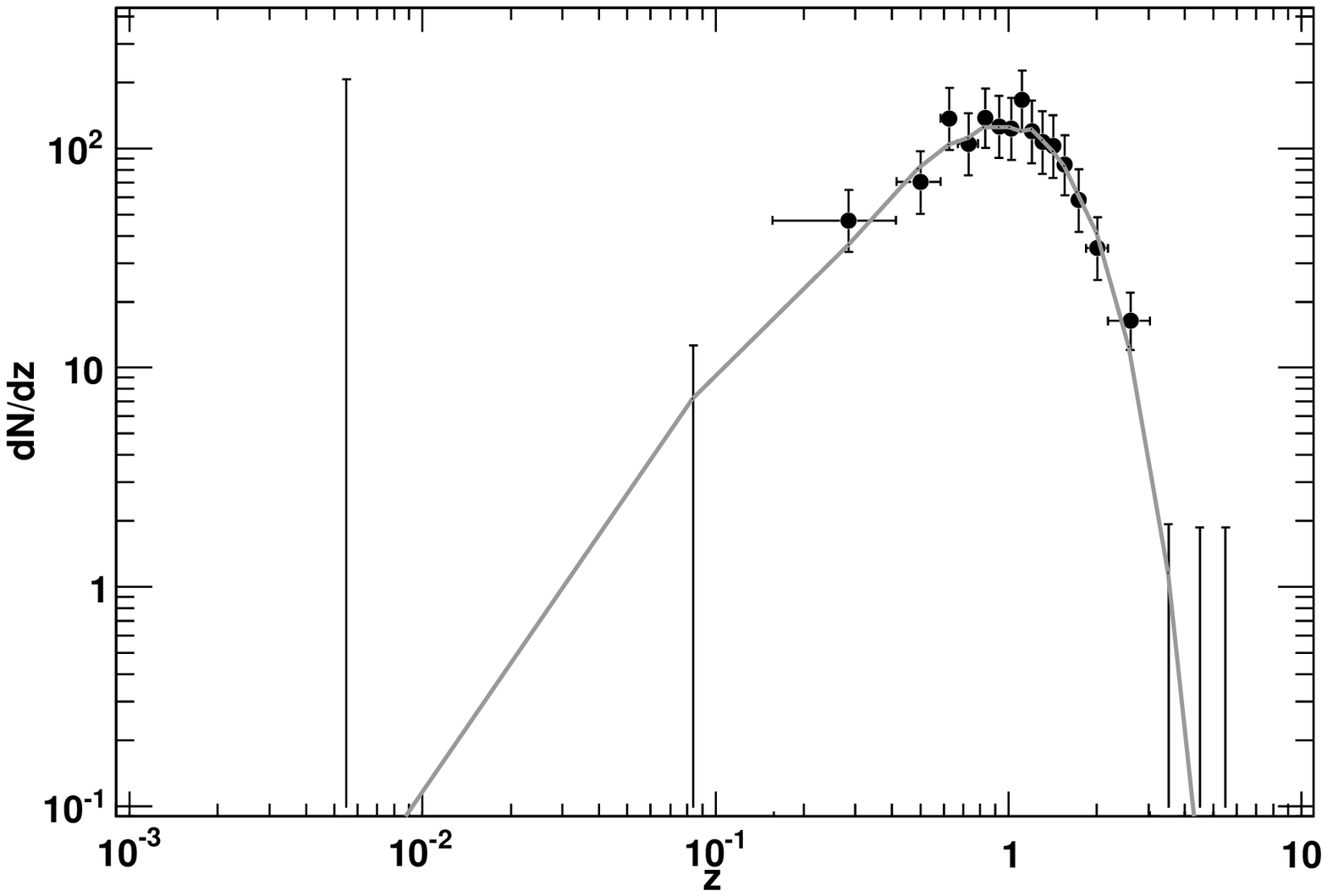} &
\hspace{-1cm}
  	 \includegraphics[scale=0.32]{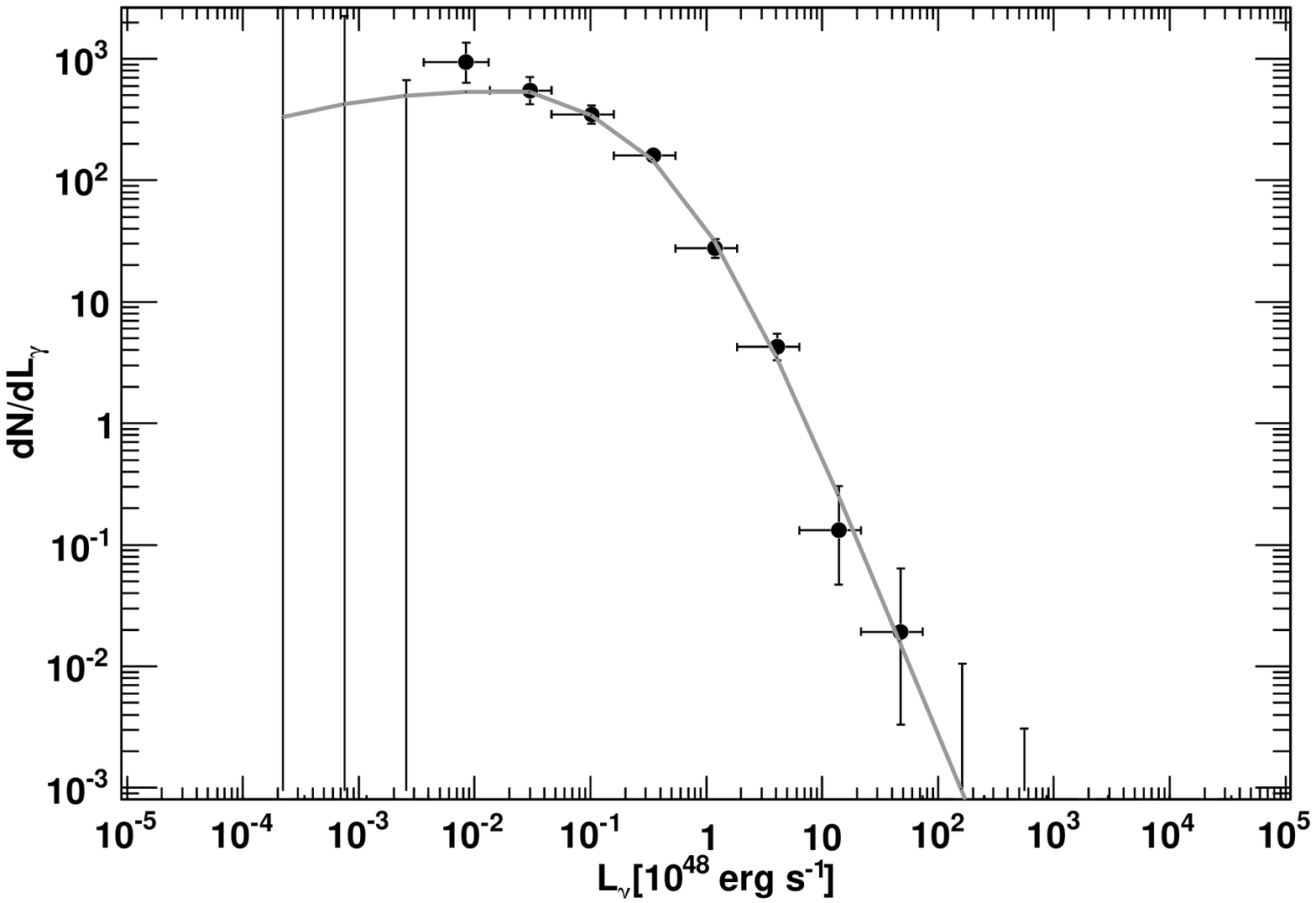} &
\hspace{-1cm}
	 \includegraphics[scale=0.32]{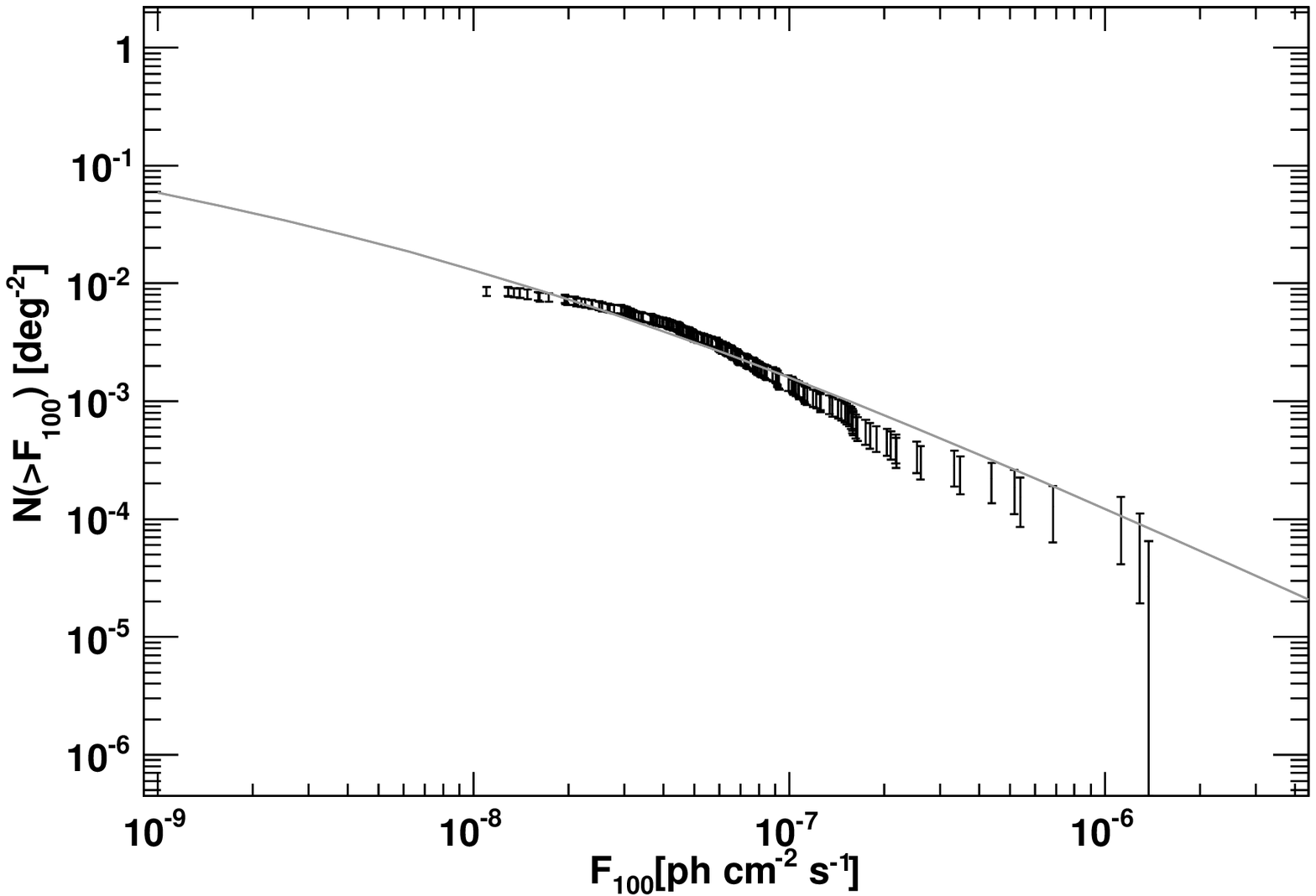}\\
\end{tabular}
  \end{center}
  \caption{Redshift (left), luminosity (middle) and source count (right)
distribution of LAT FSRQs. The dashed line is the best-fit PLE model discussed
in the text.
\label{fig:ple}}
\end{figure*}

In this model the evolution is entirely in luminosity: i.e. the FSRQ were more
luminous in the past if positive evolution ($k>0$) is found (the opposite is 
true otherwise).  It is also straightforward to demonstrate that the 
luminosity evolution (i.e. Eq.~\ref{eq:ple2})
of FSRQs peaks at $z_c$=$-1-k\xi$.  The best-fit parameters are reported in 
Table~\ref{tab:ple}.  The evolution of the FSRQ class is found to be positive and 
fast ($k=5.70\pm1.02$). The redshift peak is $z_c=1.62\pm0.03$. Moreover,
the subsequent rate of decrease of the luminosity after the peak 
is well constrained
($\xi=-0.46\pm0.01$). However, as shown in Fig.~\ref{fig:ple}, while 
this model provides a good fit to the observed redshift and luminosity distributions,
it is a very poor representation of the measured log$N$--log$S$.

\begin{deluxetable}{lccccccccc}
\tablewidth{0pt}
\tabletypesize{\scriptsize}
\tablecaption{Best-fit parameters of the Pure Luminosity Evolution LF. Parameters without an error
estimate were kept fixed during the fit.
\label{tab:ple}}
\tablehead{\colhead{Sample}  & \colhead{\# Objects} & 
\colhead{A\tablenotemark{a}} & \colhead{$\gamma_1$} & 
\colhead{L$_*$}              & \colhead{$\gamma_2$} &
\colhead{k}                  & \colhead{$\xi$} &
\colhead{$\mu$}              & \colhead{$\sigma$}
}
\startdata
ALL & 186 & 5.59($\pm0.41$)$\times10^3$ & 0.29$\pm0.53$ & 0.026$\pm0.066$ & 1.25$\pm0.32$ & 5.70$\pm1.02$ & -0.46$\pm0.01$ & 2.45$\pm0.13$ & 0.18$\pm0.01$\\

Low L & 89 & 15.4($\pm0.2$)$\times10^3$ & 0.29 & 0.026 & 1.25  & 4.30$\pm2.39$ & -0.50$\pm0.04$ & 2.47$\pm0.04$ & 0.21$\pm0.03$\\

High L & 97 & 22.6($\pm2.0$)$\times10^3$ & 0.29 & 0.026 & 1.25  & 3.47$\pm1.73$ & -0.79$\pm0.04$ & 2.46$\pm0.02$ & 0.20$\pm0.02$\\

\enddata
\tablenotetext{a}{In unit of $10^{-13}$\,Mpc$^{-3}$  erg$^{-1}$ s.}
\end{deluxetable}

We next test whether the redshift peak depends on luminosity,
splitting the data set at $L_{\gamma}=3.2\times10^{47}$\,erg s$^{-1}$.
A fit is then performed to each sub-sample to determine the position 
of the redshift peak (if any), keeping the parameters of Eq.~\ref{eq:ple1} fixed.
  The results of the fits to the low- and 
high-luminosity data sets are reported  in Table~\ref{tab:ple}. 
From Eq.~\ref{eq:ple2} and the values of $k$ and $\xi$
it is apparent that 
there is
a significant shift in the redshift peak, with the low- and high- luminosity
samples peaking at  $\sim$1.15 and $\sim$1.77, respectively.

\subsection{The Luminosity Dependent Density Evolution and the Redshift Peak}
\label{sec:ldde}

Since a simple PLE LF model provides an inadequate fit to the {\it Fermi} data 
and since there is some evidence for the evolution of the redshift peak with 
luminosity, we now fit the {\it Fermi} FSRQ set to a luminosity-dependent 
density evolution model (LDDE). Here the evolution is primarily in density
with a luminosity-dependent redshift peak. The LDDE model is parametrized
\begin{equation}
\Phi(L_{\gamma},z) = \Phi(L_{\gamma}) \times e(z,L_{\gamma})
\end{equation}
where
\begin{equation}
e(z,L_{\gamma})= \left[ 
\left( \frac{1+z}{1+z_c(L_{\gamma})}\right)^{p1} + 
\left( \frac{1+z}{1+z_c(L_{\gamma})}\right)^{p2} 	
 \right]^{-1}
\label{eq:evol}
\end{equation}
and 
\begin{equation}
z_c(L_{\gamma})= z_c^*\cdot (L_{\gamma}/10^{48})^{\alpha}  .
\label{eq:zpeak}
\end{equation}

$\Phi(L_{\gamma})$ is the same double power law used in Eq.~\ref{eq:ple1}.
This parametrization is similar to that proposed by \cite{ueda03}, but is 
continuous around the redshift peak $z_c(L_{\gamma})$. This has obvious 
advantages for fitting algorithms that rely on the derivatives of the fitting 
function to find the minimum.  Here $z_c(L_{\gamma})$ corresponds to the 
(luminosity-dependent) redshift where the evolution changes sign (positive to 
negative), with $z_c^*$ being the redshift peak for a FSRQ with a luminosity 
of 10$^{48}$\,erg s$^{-1}$.

\begin{figure*}[ht!]
  \begin{center}
  \begin{tabular}{cc}
\hspace{-1cm}
    \includegraphics[scale=0.45]{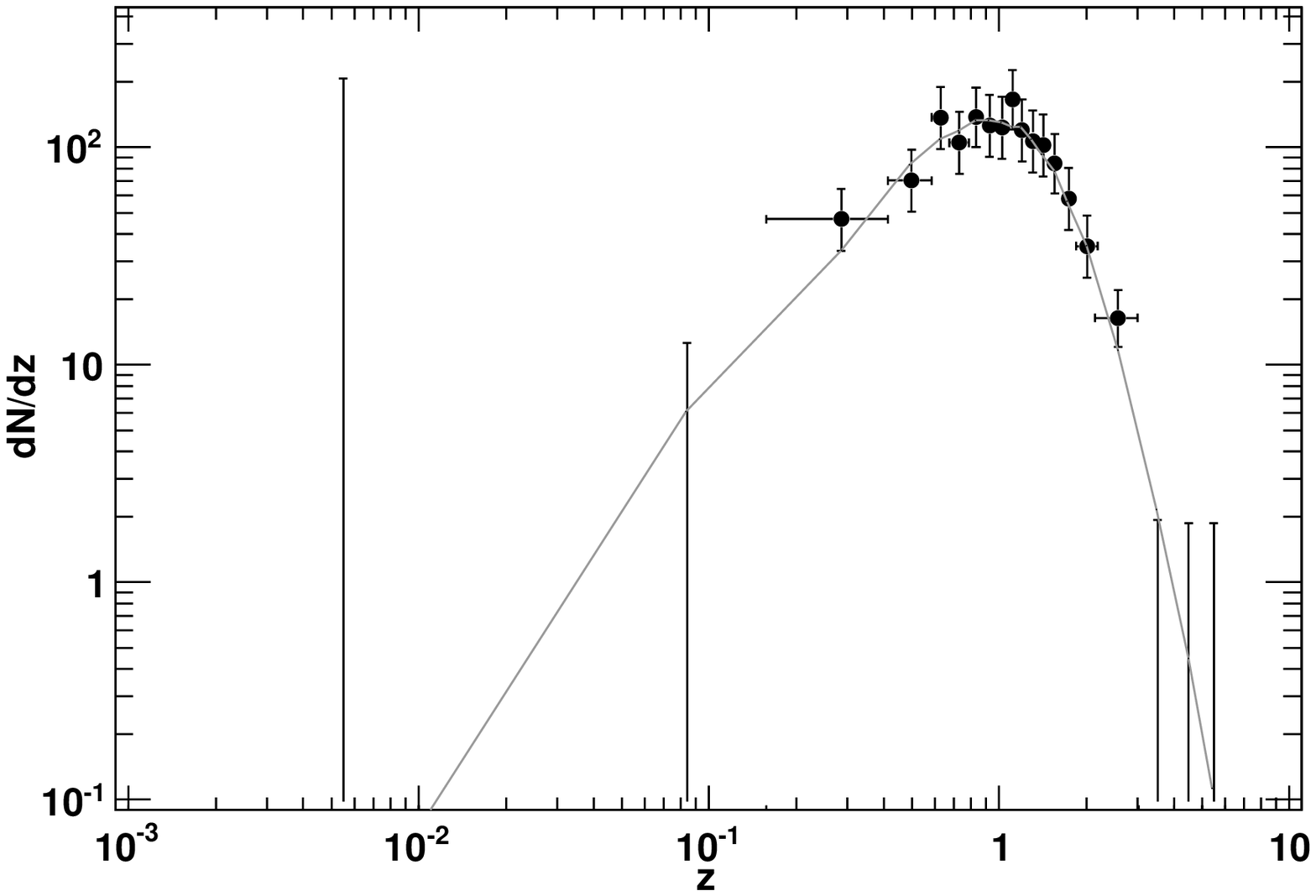} &
\hspace{-1cm}
  	 \includegraphics[scale=0.45]{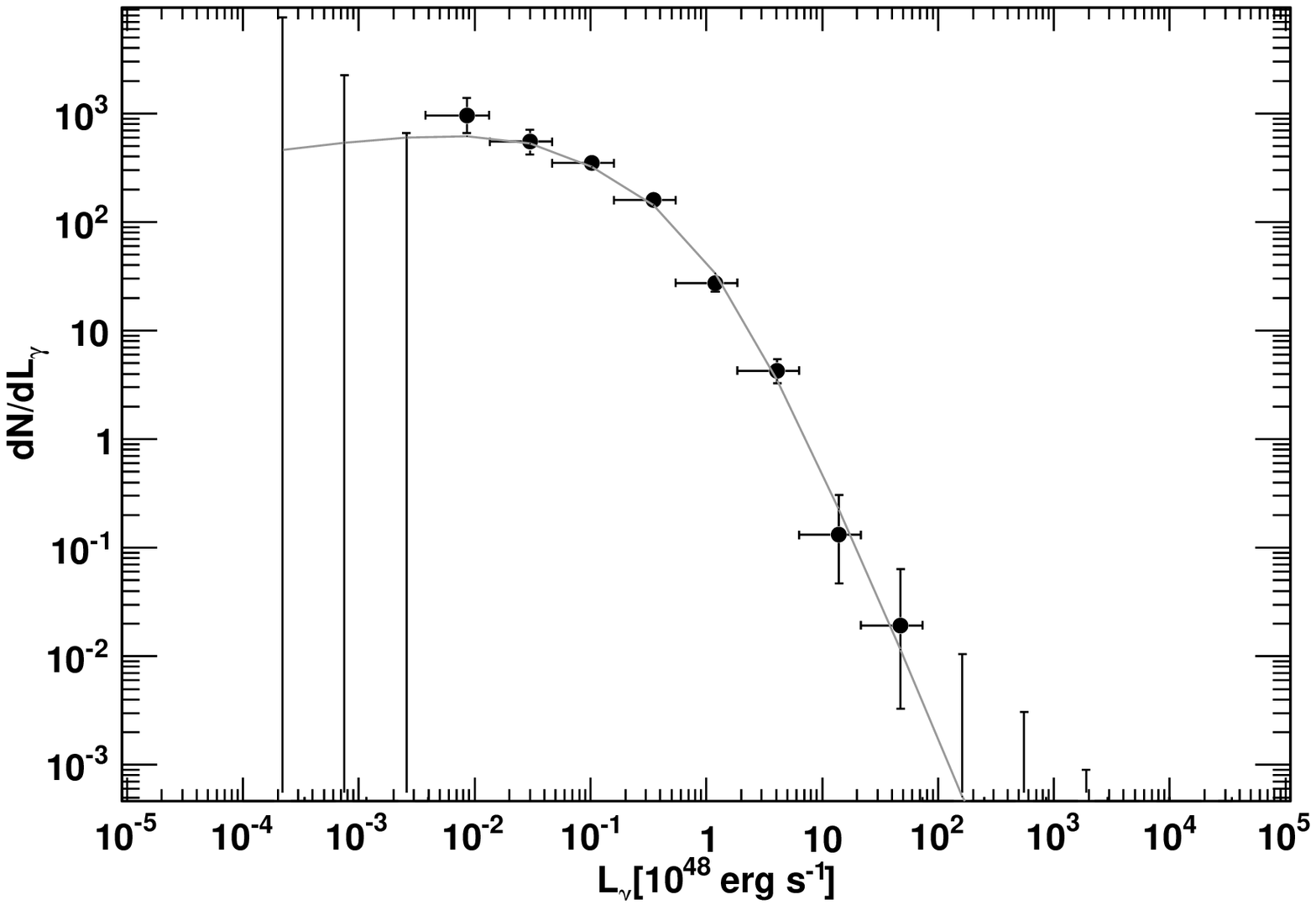} \\
\hspace{-1cm}
 	 \includegraphics[scale=0.45]{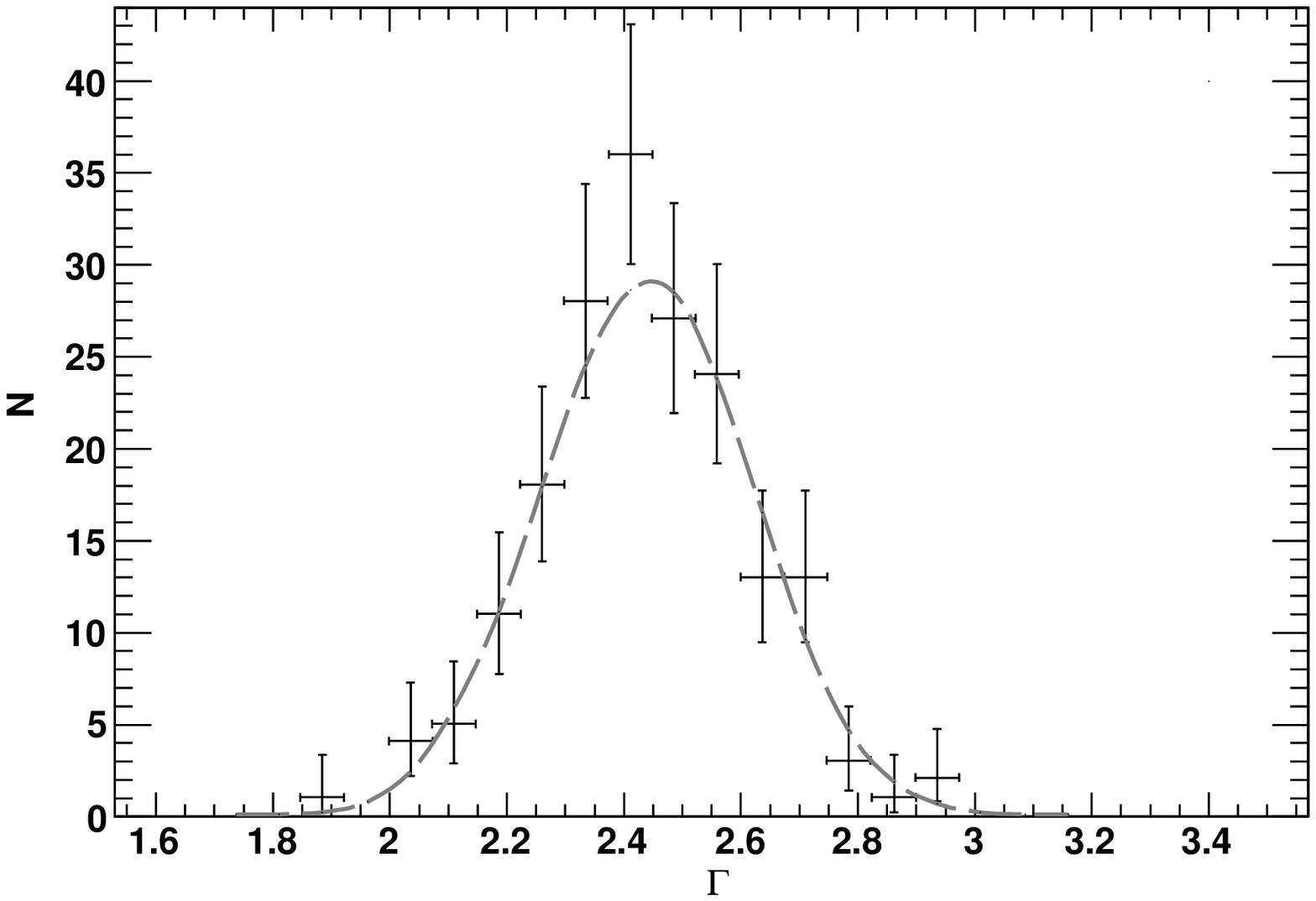} &
\hspace{-1cm}
	 \includegraphics[scale=0.45]{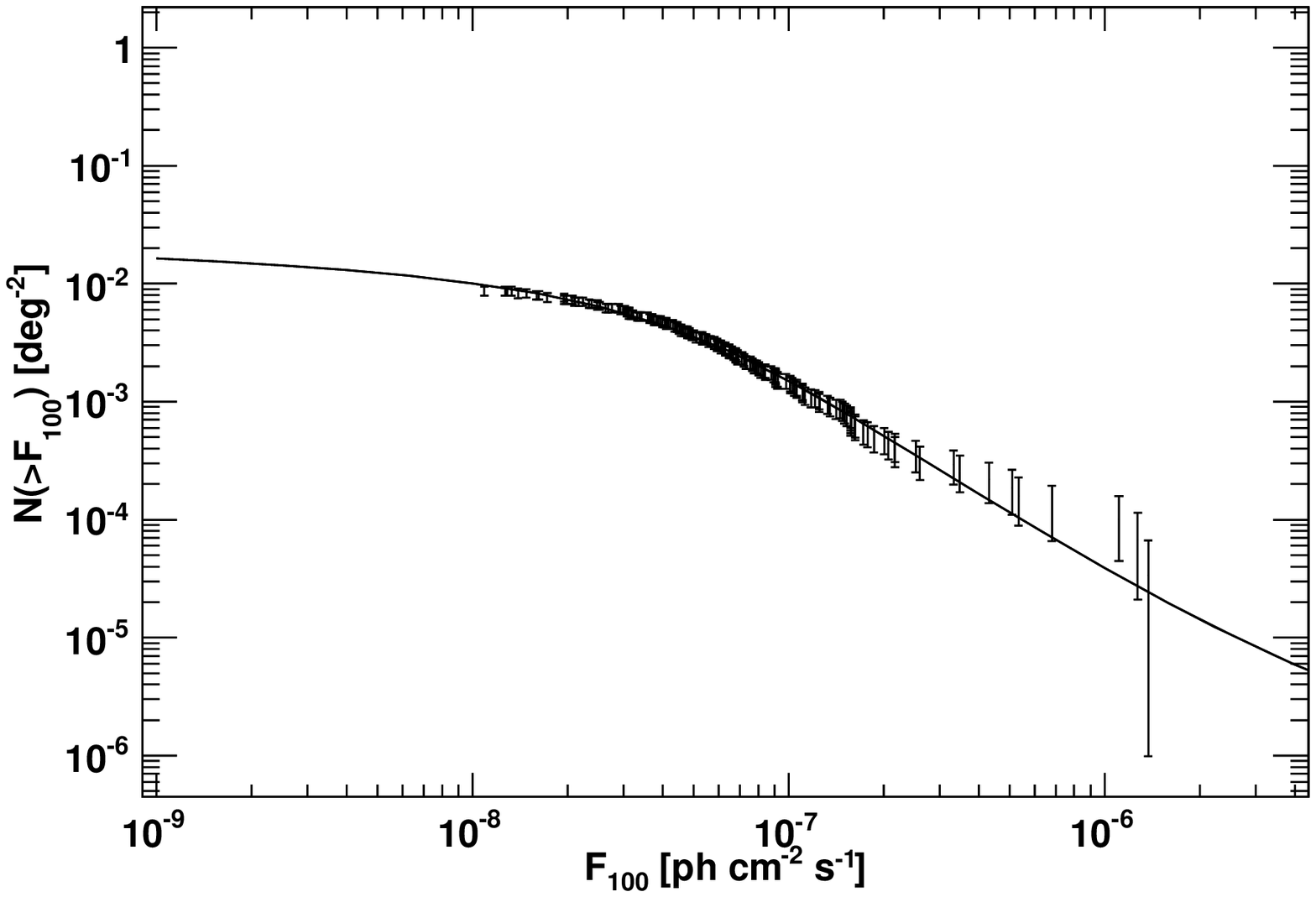}
\end{tabular}
  \end{center}
\caption{Redshift (upper left), luminosity (upper right), 
photon index (lower left),  and source count (lower right)
distributions of LAT FSRQs. The dashed line is the best-fit LDDE model convolved
with the selection effects of {\it Fermi}.
Notice the greatly improved source count distribution over the predictions 
of the PLE model of Figure~\ref{fig:ple}.
\label{fig:ldde}}
\end{figure*}

The LDDE model provides a good fit to the LAT data and is able to reproduce the observed distribution in Fig.~\ref{fig:ldde}.
The log-likelihood ratio test shows that the improvement over the best PLE model is 
significant, with a chance probability of  $\sim10^{-6}$.
Results are reported in Table ~\ref{tab:ldde}.
%
%

\begin{figure}[h!]
\begin{centering}
	\includegraphics[scale=0.9]{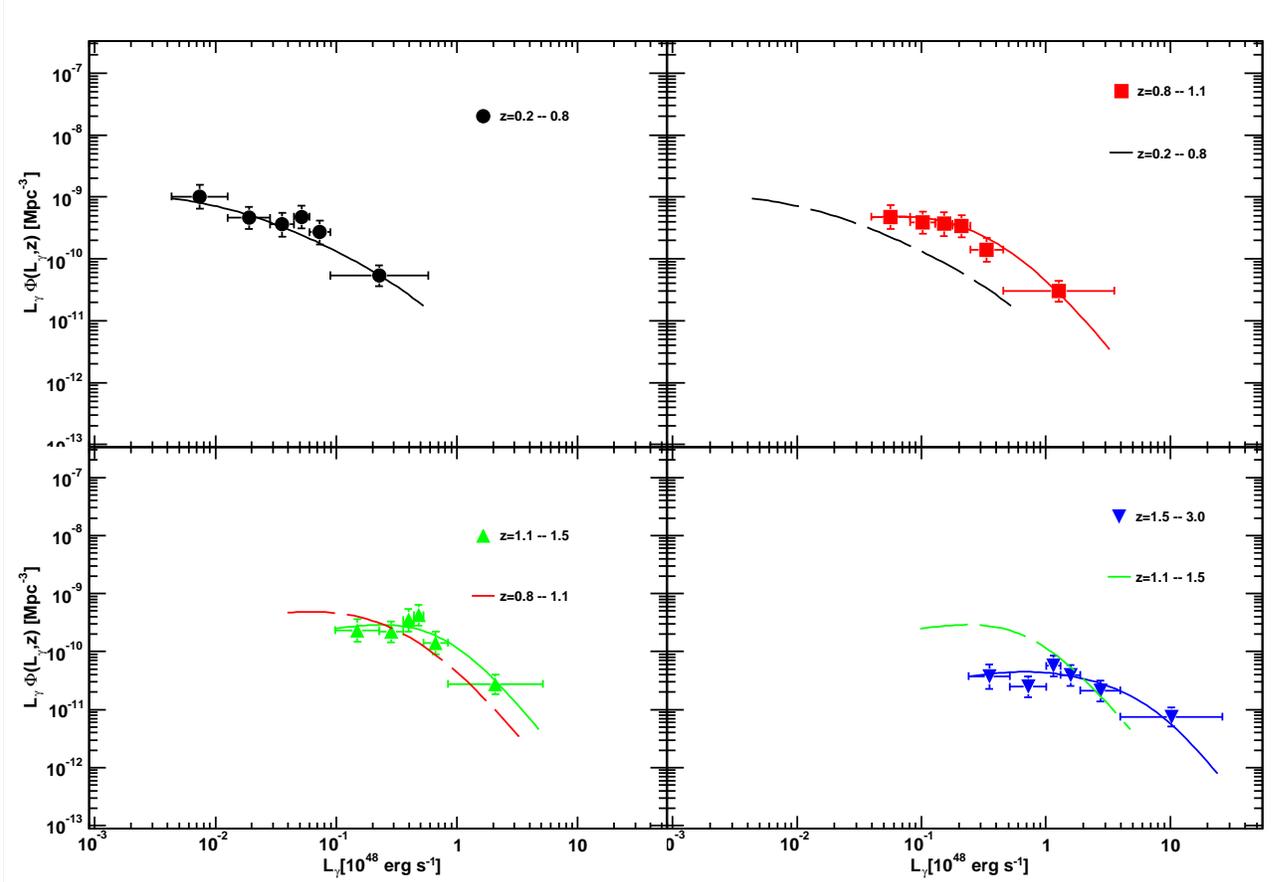} 
	\caption{LF of the {\it Fermi} FSRQs in different bins of redshift,
reconstructed using the $N_{obs}/N_{mdl}$ method. The lines represent the
best-fit LDDE model of $\S$~\ref{sec:ldde}. To highlight the evolution,
the LF from the next lower redshift bin is over-plotted (dashed lines).
	\label{fig:glf}}
\end{centering}
\end{figure}

In Fig.~\ref{fig:glf} we subdivide the sample into four redshift bins with
comparable number of sources to illustrate how the LF changes.
The evolution, visible as a shifting of the peak and a change of the shape
of the LF between different
bins of redshift, is clearly visible. This evolution takes place mostly
below redshift $\sim$1.1 where the space density of our least luminous objects
(i.e. L$_{\gamma}\approx10^{46}$\,erg s$^{-1}$) increases by $\sim 10\times$.
Above this redshift the variation is less marked, but one notices that:
\begin{itemize}
\item the space density of $\log$L=47 objects decreases from redshift 1 to 
redshift 1.5 while that of $\log$L=48 FSRQs still increases
(lower left panel, green versus red line).  This indicates that the space 
density of $log$L=47 FSRQs peaks at a redshift $\sim1.1$.
\item similar behavior 
holds for $log$L=48 FSRQs in the highest redshift bin so that their maximum
space density should occur well below z=3.
\end{itemize}

\begin{figure}[h!]
\begin{centering}
	\includegraphics[scale=0.8]{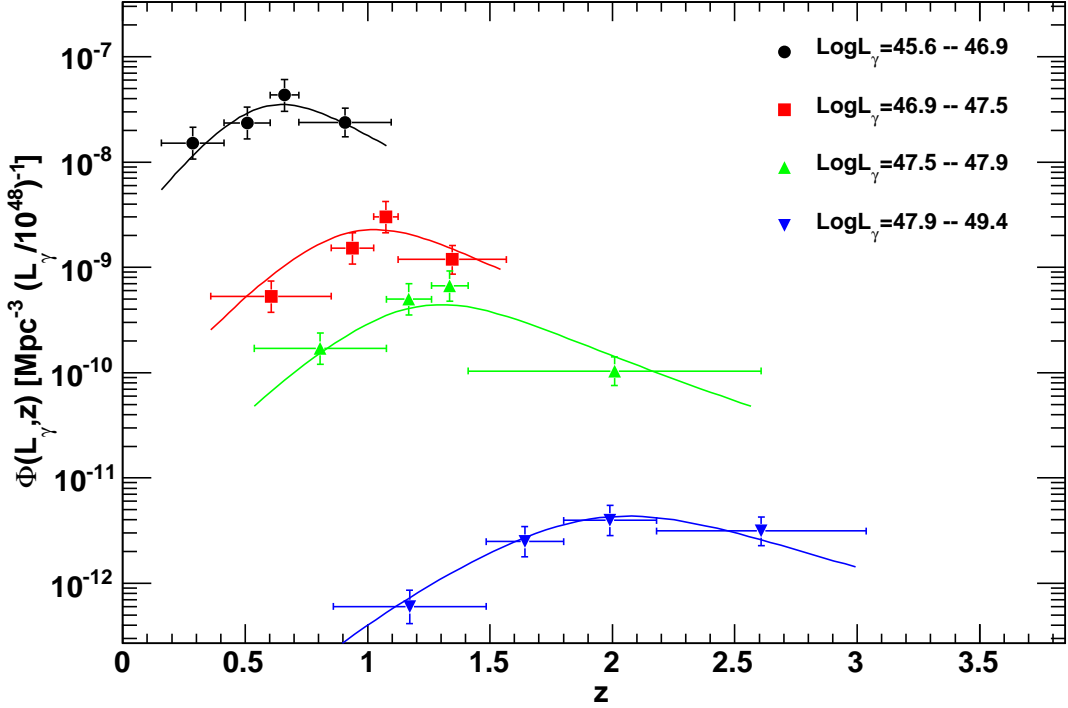} 
	\caption{Growth and evolution of different luminosity classes of 
FSRQs. Note that the space density of the most luminous FSRQs peaks
earlier in the history of the Universe while the bulk of the 
population (i.e. the low luminosity objects) are more abundant at
later times.  The range of measured distribution is determined by requiring at least one source within the volume (lower left) and sensitivity limitations of {\it Fermi} (upper right).
	\label{fig:zpeak}}
\end{centering}
\end{figure}

The best-fit parameters confirm that the redshift of maximum
space density increases with increasing luminosity (with the power-law index
of the redshift-peak evolution $\alpha=0.21\pm0.03$, see Eq.~\ref{eq:zpeak}). This redshift evolution can be
clearly seen in Fig.~\ref{fig:zpeak}, which shows the change in space density
for different luminosity classes.

\begin{deluxetable}{lccccccccccc}
\tablewidth{0pt}
\tabletypesize{\scriptsize}
\rotate
\tablecaption{Best-fit parameters of the Luminosity Dependent Density Evolution LF
\label{tab:ldde}}
\tablehead{\colhead{Sample}  & \colhead{\# Objects} & 
\colhead{A\tablenotemark{a}} & \colhead{$\gamma_1$} & 
\colhead{L$_*$}              & \colhead{$\gamma_2$} &
\colhead{z$_c^*$}            & \colhead{$\alpha$}    &
\colhead{p$_1$}              & \colhead{p$_2$}      &
\colhead{$\mu$}              & \colhead{$\sigma$}
}
\startdata
ALL & 186  & 3.06($\pm0.23$)$\times10^4$ & 0.21$\pm0.12$ & 0.84$\pm$0.49 & 1.58$\pm0.27$ & 1.47$\pm0.16$ & 0.21$\pm0.03$ & 7.35$\pm1.74$ & -6.51$\pm1.97$ & 2.44$\pm0.01$ & 0.18$\pm0.01$\\


ALL\tablenotemark{b} & 208  & 2.82($\pm0.19$)$\times10^4$ & 0.26$\pm0.12$ & 0.87$\pm$0.53 & 1.60$\pm0.27$ & 1.42$\pm0.15$ & 0.20$\pm0.03$ & 8.21$\pm1.78$ & -5.66$\pm1.73$ & 2.42$\pm0.01$ & 0.19$\pm0.01$\\

ALL\tablenotemark{c} & 186  & 8.72($\pm0.63$)$\times10^3$ & 0.38$\pm0.16$ & 0.89$\pm$0.70 & 1.60$\pm0.30$ & 1.38$\pm0.18$ & 0.18$\pm0.03$ & 7.71$\pm1.84$ & -4.44$\pm1.78$ & \nodata & \nodata\\

\enddata
\tablenotetext{a}{In unit of $10^{-13}$\,Mpc$^{-3}$  erg$^{-1}$ s.}
\tablenotetext{b}{22 unassociated sources were included in this 
sample by drawing random redshifts from the observed redshift distribution
of FSRQs.}
\tablenotetext{c}{Derived using the detection efficiency for curved reported 
in Fig.~\ref{fig:det}}

\end{deluxetable}

\subsection{Analysis of Uncertainties}
\label{sec:sys}

One of the main uncertainties of our analysis is due to the 
incompleteness of the FSRQ
sample. In Table~\ref{tab:sample}, there are 17 sources with associated 
radio counterparts lacking optical type and redshift measurements. 
A fraction of these may be FSRQs.  In addition, there are 
29 sources without any statistically associated radio counterpart. 
The lack of radio flux  means that these cannot be FSRQ similar to those
in the {\it Fermi} sample, unless position errors
have prevented radio counterpart associations. Thus even if a few of these 
sources are mis-localized, the maximum possible incompleteness 
of our FSRQ sample is on the order of 20/(186+20), i.e. $\sim$10\,\%.

The standard way to account for this incompleteness is to correct upwards
 the normalization of the LF as to reflect the {\it likely} real number 
of FSRQs (associated  or  not) in our sample. 
Considering extra information about these sources, this likely incompleteness
is even smaller than the 10\% above. First, we find that only 50\% of 
all the radio-loud identified sources in our sample are FSRQs. 
This suggests that only 8 of the 17 radio identified
sources are FSRQs, with the remainder being BL Lac objects
 and lower luminosity AGN. 
A similar argument can be made based on the $\gamma$-ray spectral index.
 The median index for FSRQs is $\Gamma=2.44$ while only 9\% of our 
BL Lac objects have such soft spectra. There are 4 such soft sources in our set of
17 radio sources. Conservatively assuming that all are FSRQs,
 we infer that twice this number, namely 8, FSRQ are in the 
radio-detected sample. Of the 29 sources without radio counterparts,
4 are Blazar `ANTI-Associations' \cite[see][for details]{lat_lbas,agn_cat}, 
where we can definitively exclude any flat-spectrum radio source
bright enough for a {\it Fermi}-type blazar. 
While most of these 29 sources lack the very deep radio observations 
to make an ANTI-Association, all but 5 have been classified as pulsar candidates, 
based on gamma-ray spectral curvature and lack of variability. 
We thus suspect that virtually  all of these sources are other 
classes of gamma-ray emitters, eg. pulsars yet to be discovered, 
starburst galaxies, etc.
Our conclusion is that the {\it likely} incompleteness is 
only 8/(186+8) =4\%. Conservatively adding a few nominally 
radio-quiet sources from erroneous LAT localizations may allow 5\%
incompleteness. This correction has been applied in $\S$~\ref{sec:ple} and
$\S$~\ref{sec:ldde}.

Moreover, we show that our results are robust even in the unlikely
case that half of the unassociated sources are FSRQs (thus
neglecting that the unassociated FSRQs occupy a definite
part of the parameter space).
In this scenario, the number 
of {\it likely} yet unidentified FSRQs is $\sim$22 (i.e. 8 + 29/2).
We assign a random redshift, drawn
from a smoothed version of the FSRQ redshift distribution
to 22 out of the 46 unassociated objects (i.e. giving 10\,\% maximum
incompleteness for the FSRQ sample).
These are then used with the associated  FSRQs to derive the luminosity
function. One example is reported in Table~\ref{tab:ldde}.
It is apparent that including the unassociated
sources with random redshifts does not modify the shape and 
the parameters of the LF. Indeed, all the parameters are 
well within the statistical uncertainties of the parameters
of the LF derived using only the associated data set.
This test shows that the systematic
uncertainty introduced by the incompleteness is smaller than
the statistical uncertainty. In Appendix ~\ref{app:sys} we present
a more detailed discussion of other sources of uncertainty.

\subsection{Comparison with Previous Results}

\subsubsection{The Local Luminosity Function}
\label{sec:local}

The local LF is the luminosity function at redshift zero.
For an evolving population, the local LF is obtained by 
de-evolving the luminosities (or the densities) according to
the best-fit model. This is generally done using the 
1/V$_{\rm MAX}$ method of \cite{schmidt68}, as reported
for example by \cite{dellaceca08b}. However, since the best representation
of the LF is the LDDE model, the maximum volume has to be weighted 
by the density evolution relative to the luminosity of the source. 
In this case, the maximum allowed volume for a given source is defined as:
\begin{equation}
V_{\rm MAX} = \int^{z_{max}}_{z_{min}} \Omega(L_i,z) \frac{e(z,L_i)}{e(z_{min},L_i)}
  \frac{dV}{dz}dz
\end{equation}
where $L_i$ is the source luminosity, $\Omega(L_i,z)$ is the sky coverage,
$z_{max}$ is the redshift above which the source drops out of the survey,
and $e(z,L_i)$ is the evolution term of Eq.~\ref{eq:evol} normalized
(through $e(z_{min},L_i)$) at the redshift $z_{min}$ to which
the LF is to be de-evolved. The LF de-evolved at $z_{min}$ ($z_{min}$=0
in this case) is built using the standard 1/V$_{\rm MAX}$ method \citep{schmidt68}.

\begin{figure}[h!]
\begin{centering}
	\includegraphics[scale=0.8]{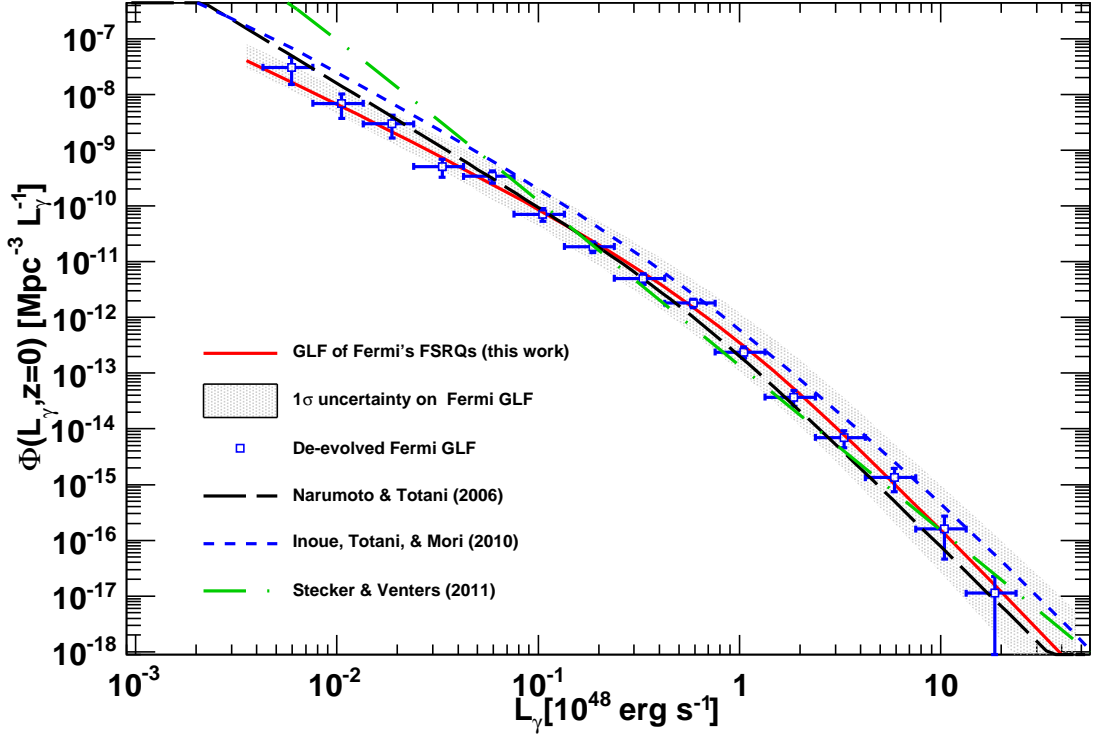} 
	\caption{Local (z=0) LF of the {\it Fermi} FSRQs as derived
from the best-fit LDDE model in $\S$~\ref{sec:ldde} (solid line).
The gray band represents the $\pm$1\,$\sigma$ uncertainty computed
as described in the text. The long- and short-dashed lines show
the LF models based on the EGRET blazars derived by 
by \cite{narumoto06} and \cite{inoue10b}respectively. The dashed-dotted line
shows the prediction from the model of \cite{stecker11}.
	\label{fig:local}}
\end{centering}
\end{figure}

In order to gauge the uncertainties that the different methods might introduce
in the determination of the local LF we consider also an alternate method.
We perform a Monte Carlo simulation, drawing 1000 series of parameters from 
the covariance matrix of the best fit LDDE model described in $\S$~\ref{sec:ldde}.
Using the covariance matrix ensures that parameters are 
drawn correctly, taking into account their correlations.
The re-sampled parameters are used to compute the $\pm1$\,$\sigma$
error of the LF at redshift zero. This is reported in Fig.~\ref{fig:local}.
There is very good agreement with the local LFs using this method and
the 1/V$_{\rm MAX}$ approach. The gray band in Fig.~\ref{fig:local} shows
the true statistical uncertainty on the space density that 
the 1/V$_{\rm MAX}$ method (applied using the best-fit parameters) 
is not able to capture.

We find a local LF described by a power law with index of 1.6--1.7, for
$F_{100}< 10^{47}$\,erg s$^{-1}$, steepening at higher luminosity.
Thus the local LF can be  parametrized as
a double power law:
\begin{equation}
\Phi(L_{\gamma}) = \frac{dN}{dL_{\gamma}}=A
\left[\left(\frac{L_{\gamma}}{L_{*}}\right)^{\gamma_1}+
\left(\frac{L_{\gamma}}{L_{*}}\right)^{\gamma2} 
\right]^{-1}
\label{eq:local}
\end{equation}
where $A=(3.99\pm0.30)\times10^{-11}$,  $L_{*}=0.22\pm0.30$, $\gamma_1$=1.68$\pm0.17$,
$\gamma_2$=3.15$\pm0.63$ and both $L_{\gamma}$ and $L_{*}$ are in
units of $10^{48}$\,erg s$^{-1}$. Other models (e.g. a Schecter
function, a simple power law etc.) do not generally provide 
as good a fit to the data. The values of the low-luminosity index $\gamma_1$
and the high-luminosity index  $\gamma_2$ are in good
agreement with that found here of 1.63$\pm0.16$ and 2.3$\pm0.3$
reported by \cite{padovani07} for the DRBXS survey of FSRQs.

Values very similar to those found here were also reported
for a radio FSRQ sample by \cite{dunlop90}, who find\footnote{Their
definition of local luminosity function and Eq.~\ref{eq:local} differ
by a 1/L$_{\gamma}$ (or 1/P in their paper) term. Thus we added 1.0 to
their exponents.} $\gamma_1=1.83$ and $\gamma_2=2.96$.  The {\it Fermi} 
LF low-luminosity index  (i.e. $\gamma_1$) is flatter than that determined 
using EGRET blazars by \cite{narumoto06} as is apparent in Fig.~\ref{fig:local}.
However, a re-analysis of the same data sets employing
the blazar sequence \citep{fossati99} to model the blazar SEDs found a 
low-luminosity index in the 1.8--2.1 range 
\citep{inoue09}. Also, in a more recent work, \cite{inoue10b} modified
their SED models to be able to reproduce TeV data of known blazars.
Their LF at redshift zero (see Fig.~\ref{fig:local}) is found to be
in relatively good agreement with that found here for the {\it Fermi} sample.

\subsubsection{The Luminosity Function at Redshift 1}

\begin{figure}[h!]
\begin{centering}
	\includegraphics[scale=0.8]{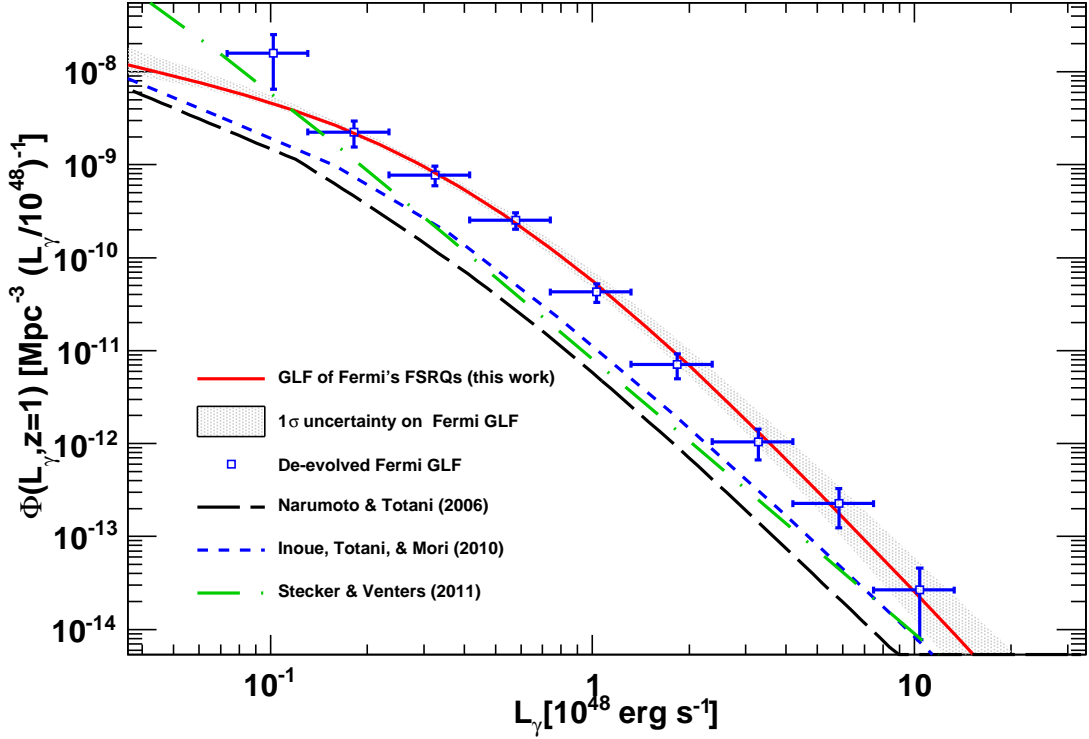} 
	\caption{LF of the {\it Fermi} FSRQs at redshift 1.0 as derived
from the best-fit LDDE model in $\S$~\ref{sec:ldde} (solid line).
The gray band represents the $\pm$1\,$\sigma$ uncertainty computed
with the method described in the text. The long dashed and short
dashed lines show
the LF models based on the EGRET blazars derived respectively
by \cite{narumoto06} and \cite{inoue10b}. The dashed-dotted line
shows the prediction from the model of \cite{stecker11}.
	\label{fig:z=1}}
\end{centering}
\end{figure}

Fig.~\ref{fig:z=1} shows the luminosity of FSRQ at redshift  1
compared to predictions from recent models. It is apparent that
the evolution of the {\it Fermi} LF is stronger than predicted
by any of these models. The increase in space density from redshift 
0 to 1 for a source with a luminosity of 10$^{48}$\,erg s$^{-1}$ 
is almost a factor $\sim150$. This dramatic increase is not seen
in the evolution of radio-quiet AGN \cite[e.g.][]{ueda03,hasinger05}
whose space density increases by a factor 25--50 between redshift 0 and 1.
The increase of a factor $\sim$60  seen in FSRQs detected in Radio
\cite{dunlop90}, is still slower than that of {\it Fermi} blazars.
This explains why the predictions based on luminosity functions derived
at other wavelengths 
(see Fig.~\ref{fig:z=1}) underpredict the density of 
high-luminosity
{\it Fermi} FSRQs at redshift of 1.

\section{The Spectral Energy Distributions of FSRQs}
\label{sec:sed}

	We may use the 0.1--300\,GeV LAT spectra of our uniform bright 
{\it Fermi} FSRQ sample along with the 15-200\,keV spectra measured by 
the {\it Swift}-BAT to  characterize the high energy Inverse Compton
(IC) sector of the blazar SED. In turn this allows us to describe
the average SED properties of the energetic blazars and to estimate
their contribution to the cosmic high-energy backgrounds.

\subsection{Data Analysis}

	For this analysis, we further restrict the sample to 
F$_{100}\geq7\times10^{-8}$\,ph cm$^{-2}$ s$^{-1}$ (corresponding
to an energy flux of 3.4$\times10^{-11}$\,erg s$^{-1}$ cm$^{-2}$ for a
power law with an index of 2.45), since for brighter
sources {\it Fermi} has a negligible bias in the detected spectral
indices and FSRQs are selected uniformly \citep{pop_pap}. For fainter
sources hard-spectrum objects (principally BL Lac objects) dominate the sample.
There are 103 bright FSRQs detected by {\it Fermi} that meet these
criteria \citep{agn_cat}.

We analyze two years of {\it Fermi} data using version V9r21 of the science
tools\footnote{http://fermi.gsfc.nasa.gov/ssc/data/analysis/software/}.
The data are filtered, removing time periods in which the instrument was not
in sky-survey mode and photons whose zenith angle is 
larger than 100\,degrees.  We consider only photons collected within 15\,degrees 
of the source position with 100\,MeV$\leq$E$\leq300$\,GeV. We employ
the P7SOURCE\_V6 instrumental response function (IRF) and perform binned 
likelihood analysis using the {\it gtlike} tool.
First, a likelihood fit using a power-law model for all the sources in the region 
of interest (ROI) is performed on the entire energy band (100\,MeV--300\,GeV).
The parameters (i.e. flux and photon index) of all the sources within 
3$^{\circ}$ of the target FSRQ, along with the normalization of the diffuse 
model, are left free. More distant sources have parameters frozen at the 2FGL
measured values \citep{2FGL}.
We next choose 30 logarithmically spaced energy bins 
 and perform a binned likelihood in each, deriving the flux of the 
target FSRQ in each energy bin. During this exercise the flux of the FSRQ is 
allowed to vary while the photon index is fixed at the best-fit 
value found for the whole band. All the neighboring sources had parameters
fixed at the best-fit values, although the diffuse emission normalization was
allowed to vary. This analysis provides a 30-bin 100\,MeV--300\,GeV energy 
spectrum for all 103 sources in the bright FSRQ sample.

{\it Swift}-BAT is a coded-mask telescope that has conducted a several year
survey in the  15--200\,keV hard X-ray sky\citep{gehrels04,barthelmy05}. With this
deep exposure, BAT reaches a sensitivity of $\sim$10$^{-11}$\,erg cm$^{-2}$ s$^{-1}$
on most of the high-latitude sky \citep[e.g.][]{tueller07,ajello08a,ajello08b,cusumano10}.
Blazars represent 15--20\,\% of the extragalactic source population
detected by BAT \citep{ajello09b}. Since {\it Fermi}-LAT and {\it Swift}-BAT 
have comparable sensitivity in their respective bands\footnote{A {\it Fermi} 
FSRQ with a photon flux of 3$\times10^{-8}$\,ph cm$^{-2}$ s$^{-1}$ in the 
100\,MeV--100\,GeV band and a power-law spectrum with an index of 2.4  
has a energy flux of 1.5$\times10^{-11}$\,erg cm$^{-2}$ s$^{-1}$.}
and since the two bands cover the bulk of the IC component, a joint study allows
an accurate characterization of the IC spectrum and the contribution to
the background.

We use $\sim$6\,years of BAT data to extract a 15--200\,keV  spectrum
for all the FSRQs 
in the {\it Fermi} sample. Spectral extraction is performed according to 
the recipes presented by \cite{ajello08b} and discussed in detail by
\cite{ajello09} and \cite{ajello10}.  Both BAT and LAT spectra are multiplied by 
4$\pi D_L(z)^{2}$ (with $D_L(z)$ the luminosity distance at redshift z) to transform
the flux into a luminosity and shifted by (1+z) to transform into source rest-frame SEDs.

	For each FSRQ, we fit the BAT and LAT data with an empirical model 
of the following form:
\begin{equation}
E^{2} \frac{dN}{dE} \cdot4\pi D_L(z)^{2}  = E^{2} \left[ 
\left(\frac{E}{E_b} \right)^{\gamma_{BAT}} +
\left(\frac{E}{E_b} \right)^{\gamma_{LAT}} \right]^{-1}
\cdot e^{-\sqrt{E/E_c}} \cdot 4\pi D_L(z)^{2}
\label{eq:sed}
\end{equation}
where $\gamma_{BAT}$ and $\gamma_{LAT}$ are the power-law indices in the BAT 
and the LAT bands and E$_{b}$ and E$_{c}$ are the break and the cut-off energy,
respectively. The $e^{-\sqrt{E/E_c}}$ term allows us to model the curvature 
that is clearly visible in a few of the {\it Fermi} spectra. The fit is performed only 
for E$<20$\,GeV to avoid possible steeping due to the absorption of 
$\gamma$-ray photons by the extra-galactic background light 
\citep[EBL; e.g.][]{stecker06,franceschini08}.

\begin{figure*}[ht!]
  \begin{center}
  \begin{tabular}{ccc}
    \includegraphics[scale=0.45]{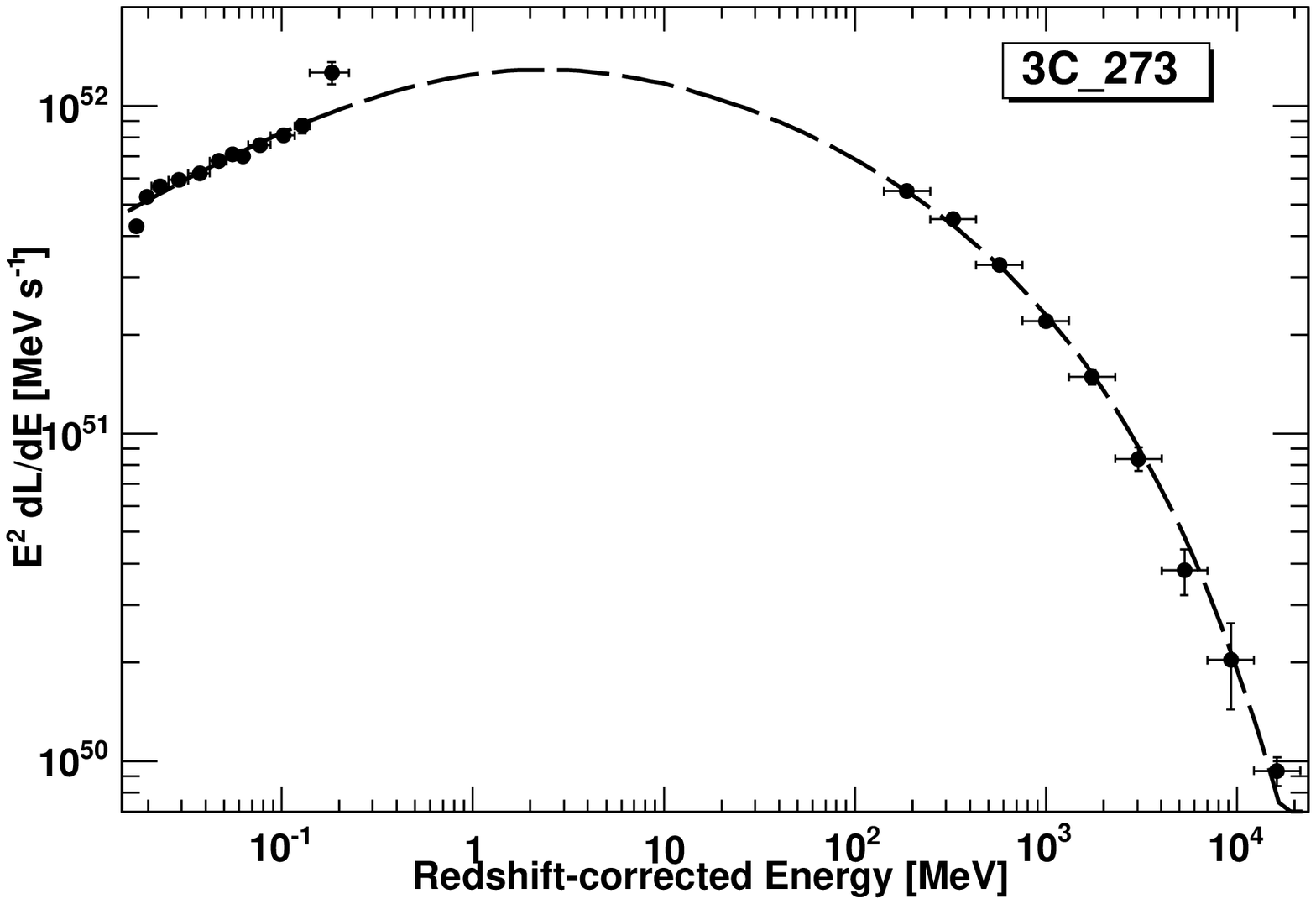} 
  	 \includegraphics[scale=0.45]{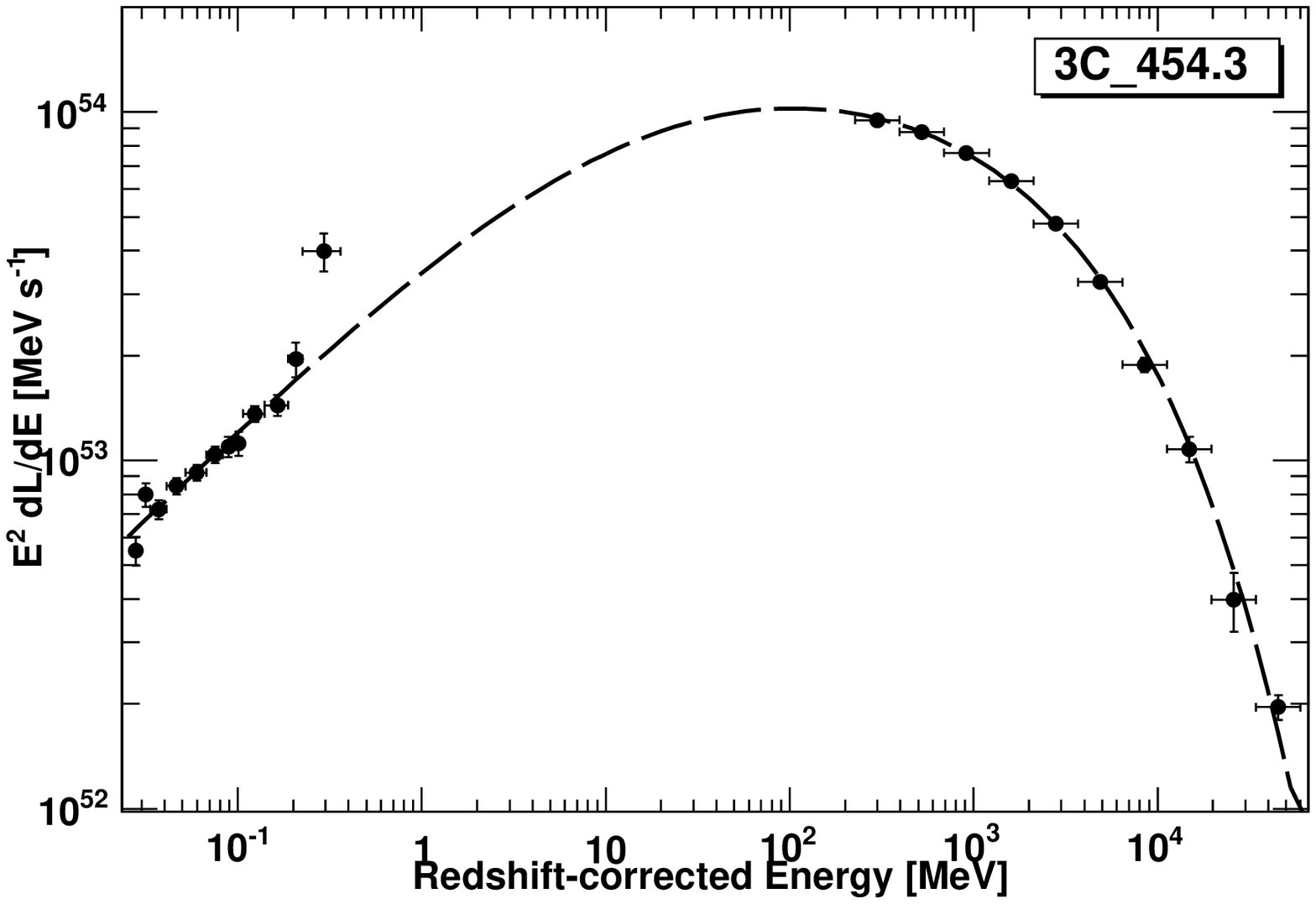} \\
\end{tabular}
  \end{center}
  \caption{Two BAT--LAT spectra of famous blazars fitted with the empirical
model described in $\S$~\ref{sec:sed}.
\label{fig:spec}}
\end{figure*}

     Two sample spectra are shown in Fig.~\ref{fig:spec}. It is apparent that 
for the brightest FSRQs, BAT and LAT are efficient in constraining the shape 
of the IC emission. Even when the BAT signal is not significant, the upper 
limit from BAT still provides useful constraints on the low energy curvature 
of the SED. In a number of bright sources (see e.g. Fig.~\ref{fig:spec})
the highest-energy datapoint in BAT at $\geq$120\,keV is seen to deviate from the
 baseline fit. This deviation is at present not significant (i.e. the reduced
$\chi^2$ of the baseline fit is already $\approx$1.0), but certainly suggestive
of a second component. Observations with INTEGRAL extending to energies $\geq$200\,keV might ascertain the nature of this feature.

	Several caveats necessarily apply to our analysis.  First, the BAT and 
LAT observations are not strictly simultaneous.  LAT spectra are accumulated 
over 2\,years while the BAT data span 6\,years.  In principle one could 
restrict the BAT data to the period spanned by the {\it Fermi} observations.
In practice this would seriously limit the BAT sensitivity, weakening 
constraints on
most of the spectra.  Second, it is possible that BAT and LAT are not sampling 
exactly the same emission component.  In particular, BAT might be dominated by 
IC emission produced by the synchrotron self Compton \cite[SSC, ][]{maraschi92}
component while the LAT may be more sensitive to External Compton 
\cite[EC,][]{dermer93} emission. Ultimately detailed SED modeling with
strictly simultaneous data would be needed to eliminate these concerns, and
such work is well beyond the scope of this paper.  Bearing these caveats in 
mind, our bright sample is nearly free of selection effects other than the 
hard flux-limit threshold applied to the {\it Fermi} data. This allows an
detailed study of the average properties of the high-energy SED of FSRQs.

\subsection{Correlation of Peak Luminosity and the Energy of the Peak}

\begin{figure*}[ht!]
  \begin{center}
  \begin{tabular}{ccc}
    \includegraphics[scale=0.45]{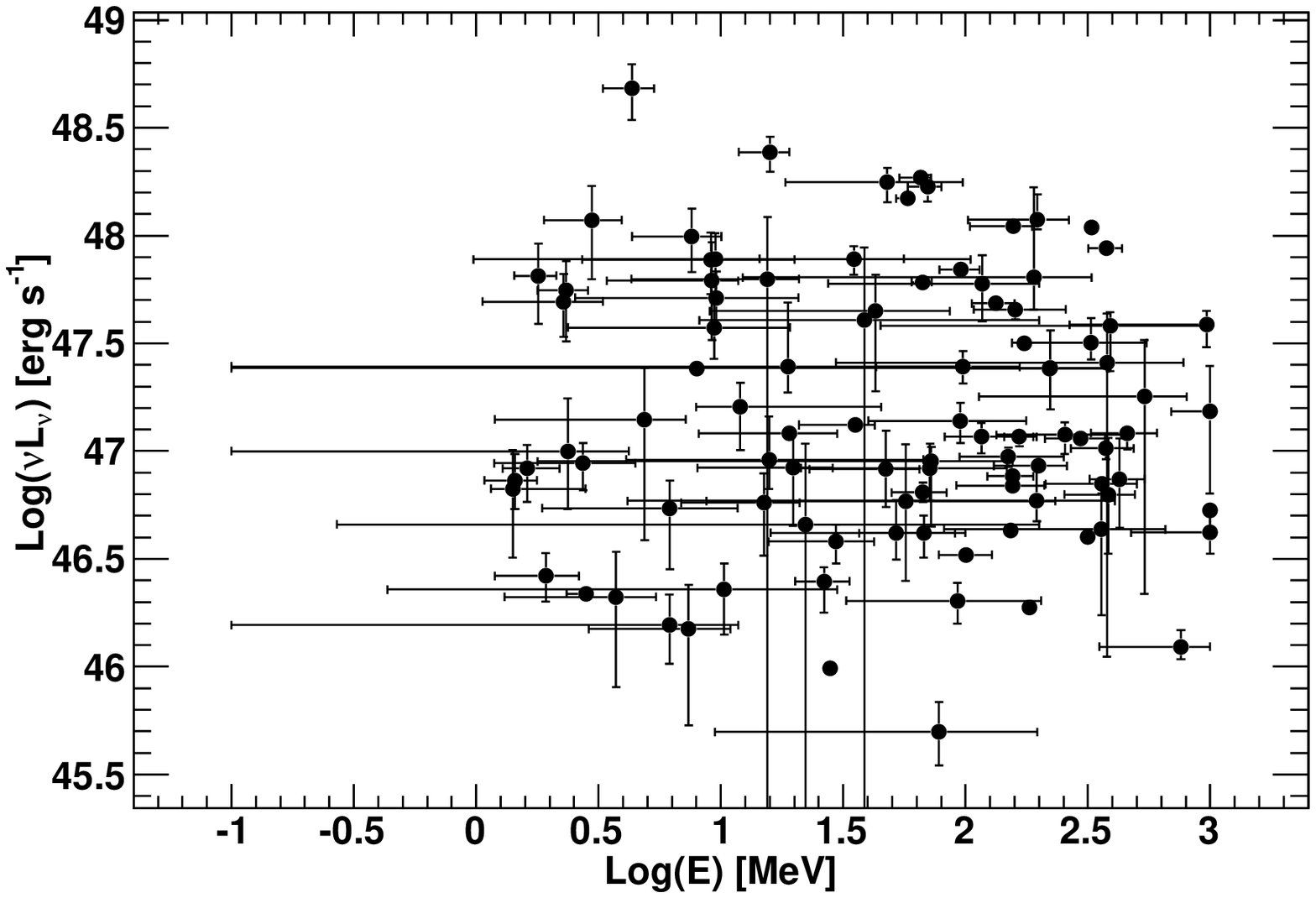} 
  	 \includegraphics[scale=0.45]{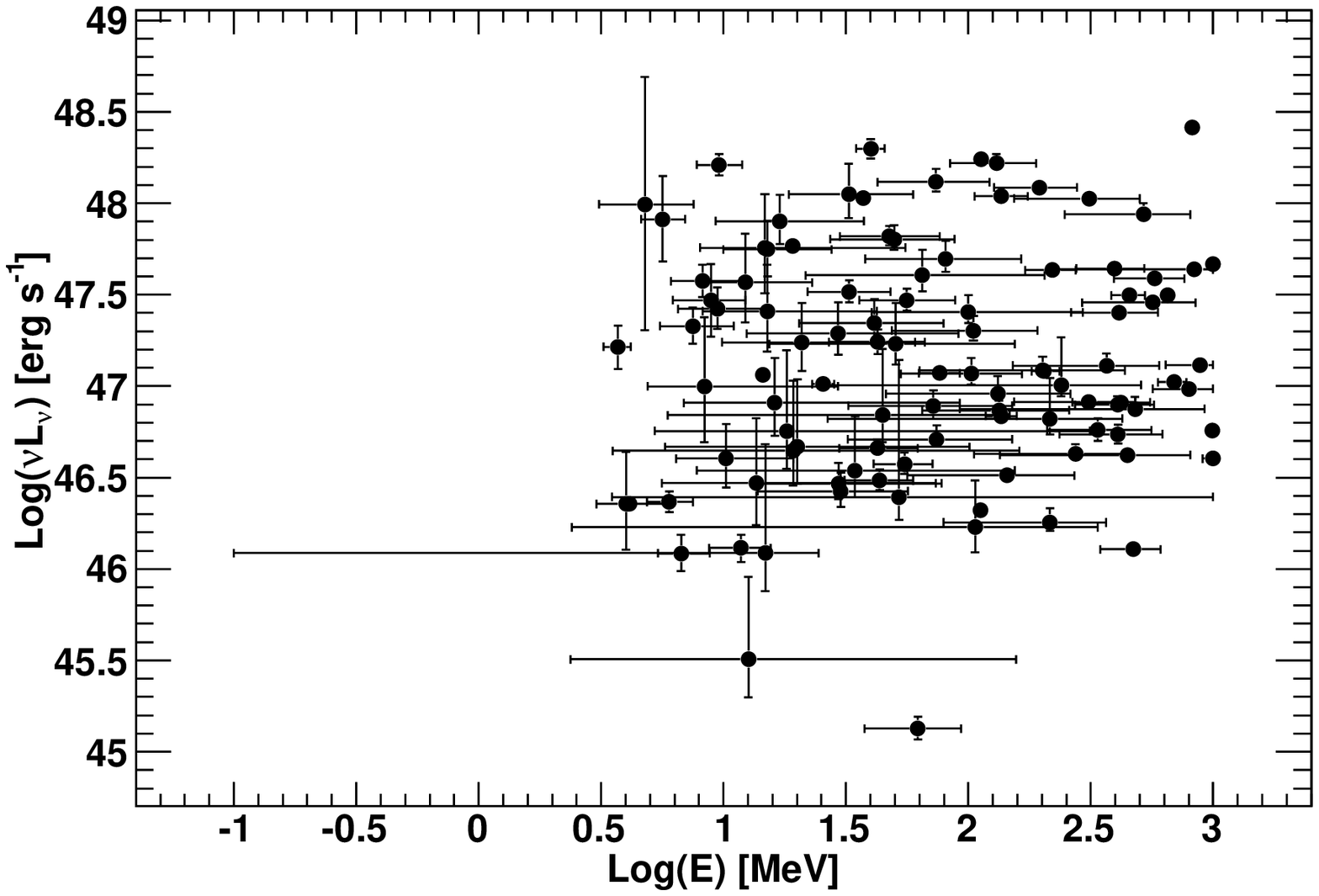} \\
\end{tabular}
  \end{center}
  \caption{Peak luminosity versus the energy of the peak for
the complete sample of FSRQs discussed in $\S$~\ref{sec:sed}.
The left plot shows the value derived using a double power law
with exponential cut-off while the right panel show parameters derived 
using a third degree polynomial function.
\label{fig:epeak}}
\end{figure*}

	We can compare the Inverse Compton rest-frame peak luminosity and peak energy
from the SED fits to the FSRQ in our sample.
As shown in Figure~\ref{fig:epeak}, there is 
no apparent correlation between these quantities; indeed the Kendall test gives
a value $\tau=0.09$ indicating no significant correlation.  Since generally
neither {\it Fermi} nor BAT directly sample the high-energy peak, we also fit 
the spectra using a third degree polynomial
function instead of the model in Eq.~\ref{eq:sed}. The results are shown
in the right panel of Fig.~\ref{fig:epeak} and confirm the previous findings.

This is in contrast to the correlation found \cite[but see also][]{nieppola08} 
between the luminosity and the energy of the {\it synchrotron} peak of blazars 
\cite[e.g.][]{ghisellini98,fossati99}.  This might imply that the jet parameters 
(e.g. Doppler factor, luminosity of the target photon field, etc.) do not depend 
on blazar GeV luminosity or redshift. This may be understood if the IC peak is
largely controlled by EC emission for these sources.

\subsection{Average SEDs}
\label{sec:average}

It is useful to estimate the average SED of FSRQs, particularly
for estimating the  contribution of FSRQs 
to the extragalactic gamma-ray background.
First we define the bolometric luminosity as the luminosity in the 10\,keV 
-- 300\,GeV band\footnote{The best fit is extrapolated from 20\,GeV to 300\,GeV.} 
and divide the sources into four bins of bolometric luminosity
with approximately the same number of objects in each bin. In these luminosity
bins we compute the average of the logarithm of the spectral luminosity at 
each energy.  Associated errors on this average spectrum are computed using 
the Jackknife technique. In this framework we neglect uncertainties due
to different level of the energy density of the extragalactic background
light which would affect mostly the high-energy part of the SED (i.e. $\geq$20\,GeV).

\begin{figure*}[ht!]
  \begin{center}
  \begin{tabular}{c}
    \includegraphics[scale=0.8]{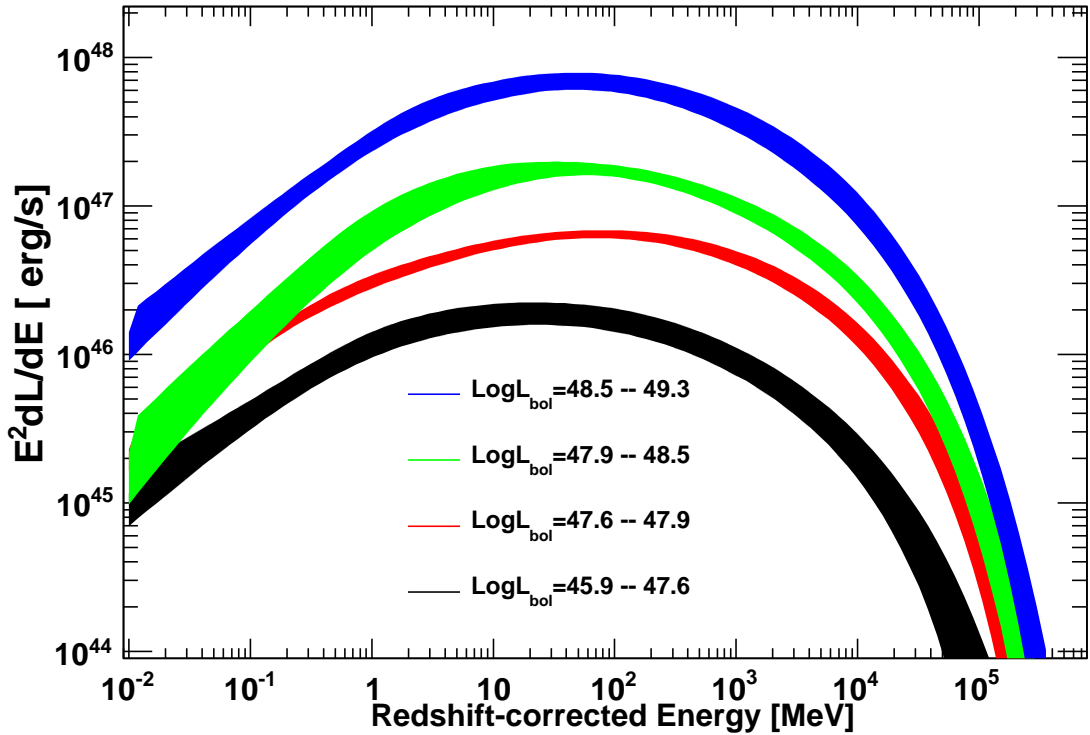} 
\end{tabular}
  \end{center}
  \caption{Average rest-frame spectral energy distributions for four 
representative FSRQ luminosity classes (left panel), see $\S$~\ref{sec:average}. 
In each SED, the band represents the 1\,$\sigma$ uncertainty on the average.
This does not reflect the uncertainty connected to the level of the extragalactic background light.
\label{fig:averagesed}}
\end{figure*}

Fig.~\ref{fig:averagesed} shows the average rest-frame SED for the FSRQ sample
in the four luminosity bins. This plot confirms the lack of a systematic
correlation of the peak luminosity and energy.  Indeed, all the averaged SEDs show
a peak in the 10--100\,MeV band and their shape
 does not change much with luminosity.

\begin{figure}[h!]
\begin{centering}
	\includegraphics[scale=0.8]{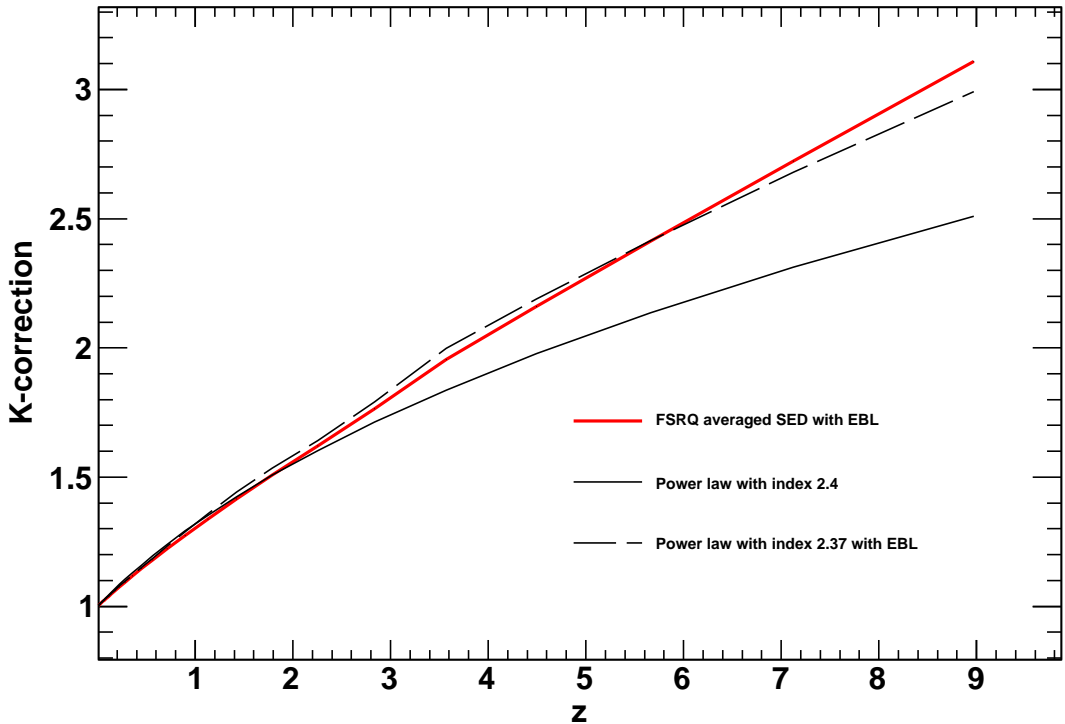} 
	\caption{Ratio of source rest-frame luminosity to observed luminosity (i.e.
k-correction) as a function of redshift for the average SED shape reported in
$\S$~\ref{sec:average} and for two generic power laws.
	\label{fig:kcorr}
}
\end{centering}
\end{figure}

	To transform luminosities between observed and rest-frame we need
the k-correction, along with its redshift variation, shown in Fig.~\ref{fig:kcorr}.
In practice, there is little difference between the k-correction for the average SED
computed here \cite[even applying EBL absorption, e.g.][]{franceschini08}
and one computed for a power law (i.e. $(1+z)^{\Gamma-2}$)
with a photon index of 2.4. Only at large redshifts do the two k-corrections
start to differ; this difference is only $\sim$5\,\% at a redshift
of 4. We find that using  a power law index of 2.37 and taking into account
EBL absorption allows us to reproduce correctly the k-correction up to redshift 
$\sim$6.

\section{The Contribution to the Isotropic Gamma-Ray Background}
\label{sec:egb}

The nature of the diffuse gamma-ray background at GeV energies remains one of 
the most interesting problems in astrophysics.  The presence of an isotropic 
component was first determined by the OSO-3 satellite \citep{kraushaar72} and 
confirmed by SAS-2 and EGRET \cite[respectively][]{fichtel95,sreekumar98}.
This isotropic component is normally referred to as the isotropic
gamma-ray background (IGRB).
{\it Fermi} recently provided a refined measurement of this isotropic component
showing that it can be adequately described as a 
single power law with an index of 2.4 in the 200\,MeV -- 100\,GeV energy range \citep{lat_edb}.

Blazars, representing the most numerous identified populations by EGRET and 
{\it Fermi} extragalactic skies, are expected to produce a substantial fraction 
of the IGRB. Typical predictions range from 20--30\,\% \citep{chiang98,muecke00,narumoto06,dermer07,inoue09} to 100\,\% \citep{stecker96,stecker11}.  Analysis of the source-count distribution showed 
that for F$_{100}\geq10^{-9}$\,ph cm$^{-2}$ s$^{-1}$ the contribution of 
{\it unresolved} blazars to the IGRB is $\sim$16\,\% in the 100\,MeV -- 100\,GeV 
band \citep{pop_pap}. Since the source counts distribution show a strong
break at a flux of $F_{100}\approx 6\times10^{-8}$\,ph cm$^{-2}$ s$^{-1}$,
it was concluded that extrapolating the source counts to zero flux would produce 
$\sim$23\,\% of the IGRB.

	Now, with a measured LF we can more robustly evaluate the emission
arising from faint FSRQs. In addition, the FSRQ SED shape study of the previous
section also allows improvement over the simple power-law type spectra assumed
by \cite{pop_pap}.

	The contribution of `unresolved' FSRQs to the IGRB can be estimated as:
\begin{equation}
F_{EGB} = \int_{z_{min}}^{z_{max}} dz \frac{dV}{dz} 
\int^{\Gamma_{max}}_{\Gamma_{min}} d\Gamma
\int^{L_{{\gamma},max}}_{L_{{\gamma},min}}
dL_{\gamma} F_{\gamma}(L_{\gamma},z) \frac{d^3 N}{dL_{\gamma}dzd\Gamma}
\left(1-\frac{\Omega(\Gamma,F_{\gamma})}{\Omega_{max}} \right)
\label{eq:cxb}
\end{equation}
where the limit of integration are the same as those of Eq.~\ref{eq:s} and 
$F_{\gamma}(L_{\gamma},z)$ is the flux of a source with rest-frame luminosity 
$L_{\gamma}$ at redshift $z$. Since we are interested in the diffuse flux
not yet resolved by {\it Fermi} \citep{lat_edb} the term 
(1-$\Omega(\Gamma,F_{\gamma})/\Omega_{max}$) takes into account the photon 
index and source flux dependence of the LAT source detection threshold
\cite[see][ for more details]{pop_pap}. The limits of integration
of Eq.~\ref{eq:cxb} are the same as those of Eq.~\ref{eq:s}.

In Eq.~\ref{eq:cxb} setting $\Omega(\Gamma,F_{\gamma})/\Omega_{max}$=0
allows us to compute the total $\gamma$-ray emission arising from the FSRQ class.
The result is 
3.13$^{+0.37}_{-0.25}\times10^{-6}$\,ph cm$^{-2}$ s$^{-1}$ sr$^{-1}$ 
in the 
100\,MeV -- 100\,GeV band.  This value should be compared with
the total sky intensity of 1.44$\times10^{-5}$\,ph cm$^{-2}$ s$^{-1}$ sr$^{-1}$,
which includes IGRB plus detected sources \citep{lat_edb}.
 Thus FSRQs make 
21.7$^{+2.5}_{-1.7}$\,\%
of this total  intensity.

\begin{figure*}[ht!]
  \begin{center}
  \begin{tabular}{ccc}
    \includegraphics[scale=0.45]{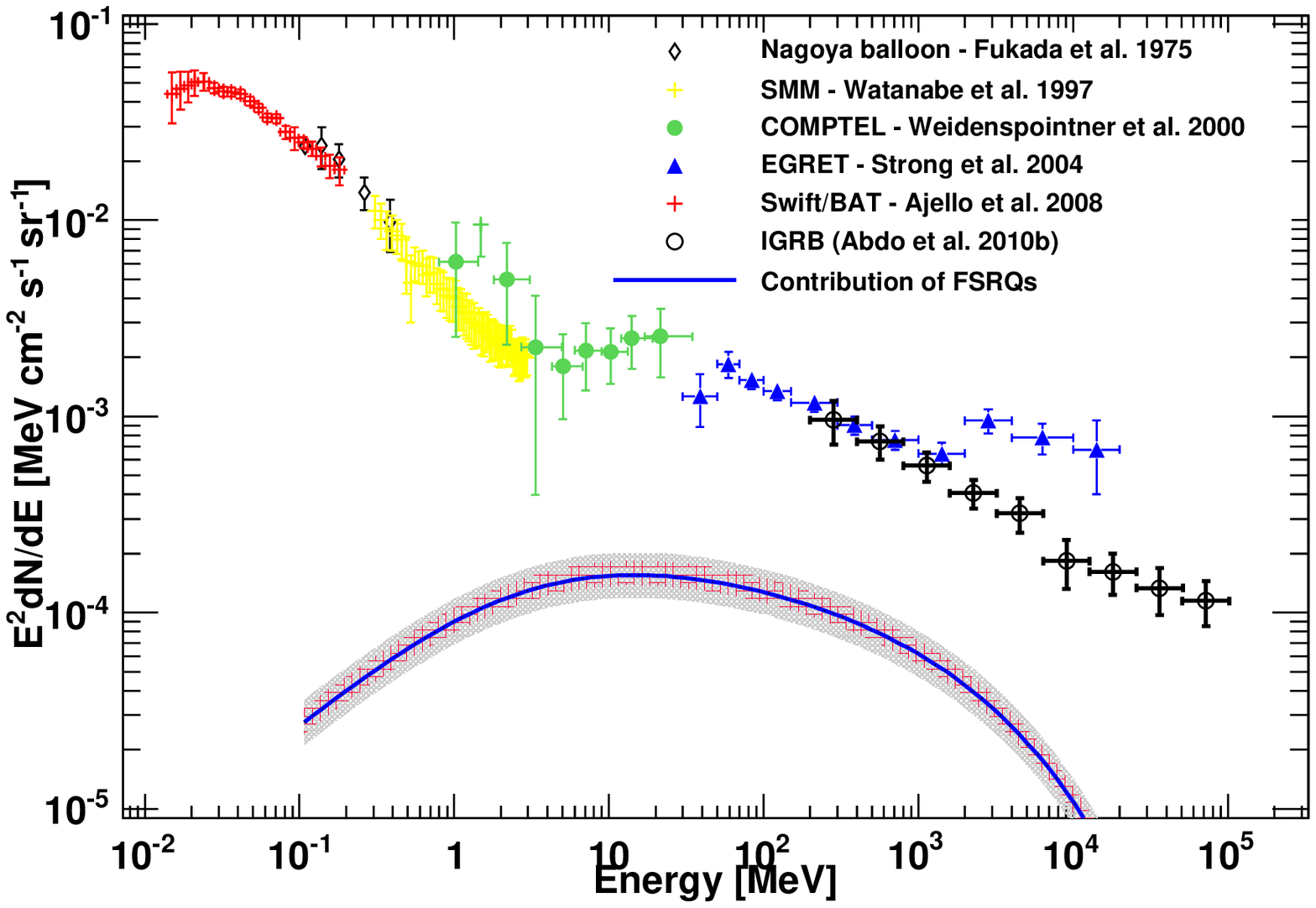}
  	 \includegraphics[scale=0.45]{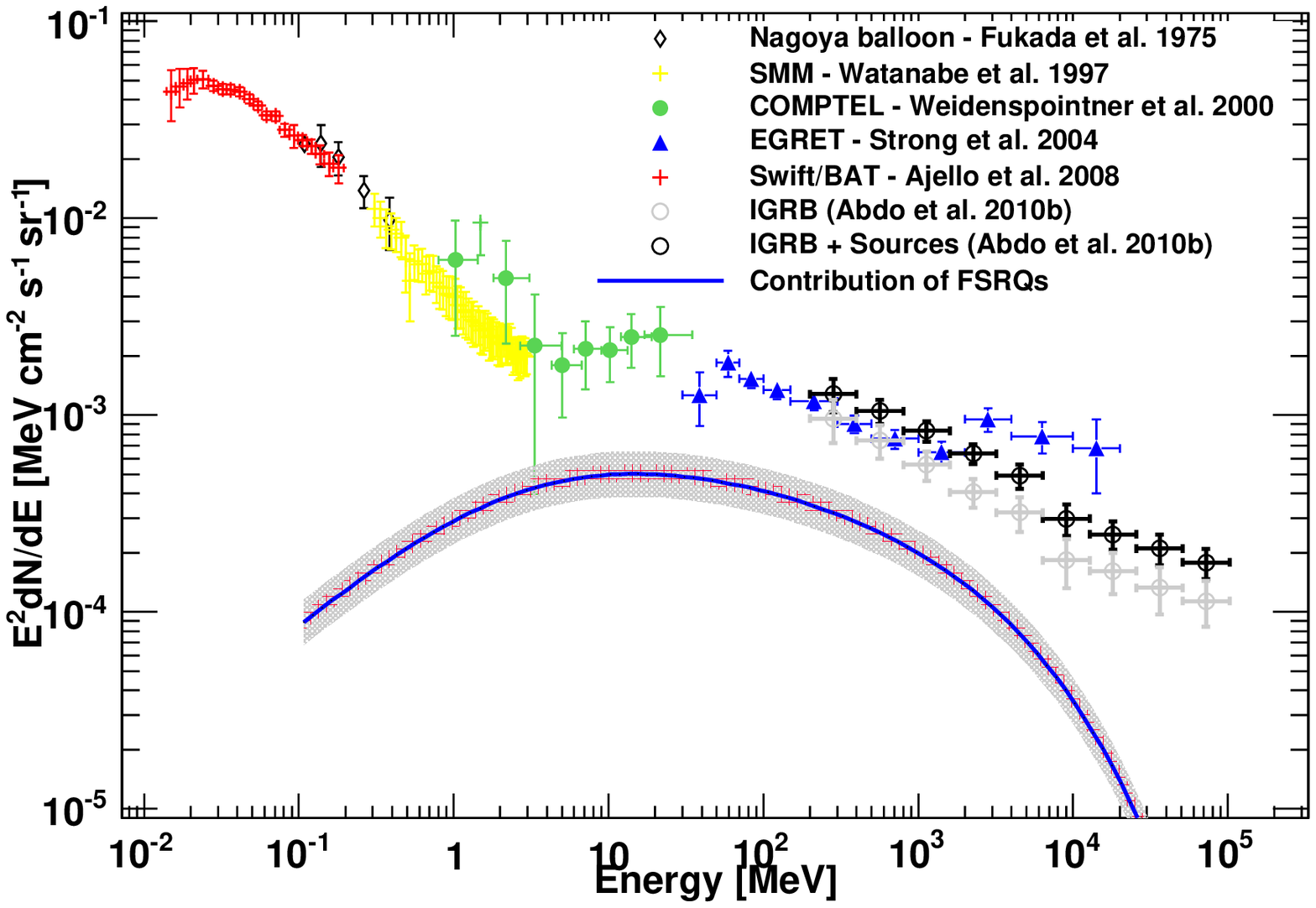} \\
\end{tabular}
  \end{center}
  \caption{Contribution of {\it unresolved} (left) and all (right) FSRQs to the 
diffuse extragalactic background (blue line) as determined by integrating the 
luminosity function coupled to the SED model derived in $\S$~\ref{sec:average}. 
The hatched band around the best-fit prediction shows the 1\,$\sigma$ statistical 
uncertainty while the gray band represents the systematic uncertainty.
	\label{fig:egb}}
\end{figure*}

If one considers the contribution only from the FSRQs
that {\it Fermi} has not detected then this becomes 
9.66$^{+1.67}_{-1.09}\times10^{-7}$\,ph cm$^{-2}$ s$^{-1}$ sr$^{-1}$
(with a maximum systematic uncertainty of 3$\times10^{-7}$\,ph cm$^{-2}$ s$^{-1}$ sr$^{-1}$ see $\S$~\ref{app:sys}).  This represents  9.3$^{+1.6}_{-1.0}$\,\%
of the IGRB intensity in the 0.1--100\,GeV band \citep{lat_edb}.
From above it is also clear that
 {\it Fermi} has already resolved more than 50\,\% of the total flux arising 
from the FSRQ class. Fig.~\ref{fig:egb} shows this contribution. The possible 
presence of external Compton components in the SEDs of FSRQs makes the estimate
between 200\,keV and 100\,MeV uncertain (see $\S$~\ref{sec:average}).
Future observations with both {\it Fermi} above 20\,MeV and INTEGRAL above
200\,keV and physical modeling of blazar spectra might
substantially reduce this uncertainty.

Even the (disfavored) 
PLE model cannot accommodate a much larger contribution of FSRQs to the IGRB.
Indeed, in this case the contribution of {\it unresolved} FSRQs would
be 1.37$\times10^{-6}$\,ph cm$^{-2}$ s$^{-1}$ sr$^{-1}$ (or $\sim$13\,\%
of the IGRB intensity).

\section{Beaming: The Intrinsic Luminosity Function and the Parent Population}
\label{sec:beaming}

The luminosities $L$ defined in this work are apparent isotropic luminosities.
Since the jet material is moving at relativistic speed ($\gamma>$1), the observed,
Doppler boosted, luminosities are related to the intrinsic values by:
\begin{equation}
L = \delta^p \mathscr{L}
\label{eq:int}
\end{equation}
where $\mathscr{L}$ is the intrinsic (unbeamed) luminosity and 
$\delta$ is the kinematic Doppler factor 
\begin{equation}
\delta = \left( \gamma -\sqrt{\gamma^2 -1}\cos\ \theta  \right)^{-1}
\end{equation}
where $\gamma=(1-\beta^2)^{-1/2}$ is the Lorentz factor and $\beta=v/c$ is the 
velocity of the emitting plasma. Assuming that the sources have a Lorentz factor 
$\gamma$ in the $\gamma_1\leq\gamma\leq\gamma_2$ range then the minimum Doppler 
factor is $\delta_{min}=\gamma_2^{-1}$ (when $\theta$=90$^{\circ}$) and the 
maximum is $\delta_{max}=\gamma_2+\sqrt{\gamma_2^2 -1}$ (when $\theta=0^{\circ}$).
We  adopt a value of  $p=4$ that applies to the case  of
continuous jet emission which seems appropriate 
for the study presented
here since  long-term average luminosities are used.
The case  $p=4$ applies also to spherical blobs if the observed 
emission is dominated by the SSC component, while
a value of $p$=5--6 should be adopted if the emission is due
to external Compton \citep{dermer95}.
However, as shown later these values imply extremely small isotropic
rest-frame luminosities.

	Beaming is known to alter the shape of the intrinsic luminosity function. 
\cite{urry84} provide an analytic solution to the case where the intrinsic 
luminosity function is a single power law and the jets have a single Lorentz 
factor. In \cite{urry91} the intrinsic luminosity function may be a double power 
law and a distribution of Lorentz factor is considered.

In this Section we will determine the intrinsic luminosity function
of the {\it Fermi} FSRQs and their Lorentz  and Doppler
factor distributions. In what follows we adopt the formalism and symbols
of \cite{lister03} and \cite{cara08}.

We begin by defining the intrinsic luminosity function as:
\begin{equation}
\Phi(\mathscr{L}) = k_{1} \mathscr{L}^{-B}
\end{equation}
valid in the $\mathscr{L}_1\leq \mathscr{L}\leq \mathscr{L}_2$ range.
The probability of observing a  beamed luminosity $L$ given a Doppler
factor $\delta$ is \citep[see also][]{lister03}:
\begin{equation}
P(L,\delta) =  P_{\delta}(\delta) * \Phi(\mathscr{L}) \frac{d\mathscr{L}}{dL}
\end{equation}
where $P_{\delta}(\delta)$ is the probability density for
the Doppler $\delta$. Assuming a random distribution for the jet angles 
(i.e. $P_{\theta}=\sin\ \theta$), 
this results in 
\begin{equation}
P_{\delta}(\delta) = \int P_{\gamma}(\gamma) P_{\theta}(\theta) \left|\frac{d\theta}{d\delta}\right| d\gamma = \int P_{\gamma}(\gamma) \frac{1}{\gamma \delta^2 \beta}d\gamma.
\end{equation}
From here it follows that
\begin{equation}
P_{\delta}(\delta) = \delta^{-2} \int^{\gamma_2}_{f(\delta)}
\frac{P_{\gamma}(\gamma)}{\sqrt{\gamma^2-1}}\ d\gamma
\end{equation}
where $P_{\gamma}(\gamma)$ is the probability density for $\gamma$ and
the lower limit of integration $f(\delta)$ depends on the Doppler
factor value and is reported in Eq.~A6 in \cite{lister03}.
Integrating over $\delta$ yields the observed luminosity function
of the Doppler beamed FSRQs:
\begin{equation}
\Phi(L) = k_1 L^{-B} \int^{\delta_2(L)}_{\delta_1(L)} P_{\delta}(\delta) \delta^{p(B-1)}
\label{eq:unbeamed}
\end{equation}
where, as in \cite{cara08}, the limits of integration are
\begin{eqnarray}
\delta_1(L) = {\rm min} \{  \delta_{max},{\rm max}\left(\delta_{min},(L/\mathscr{L}_2)^{1/p}\right) \} \\
\delta_2(L) = {\rm max} \{   \delta_{min},{\rm min}\left(\delta_{max},(L/\mathscr{L}_1)^{1/p}\right) \} 
\end{eqnarray}
In this way, by fitting Eq.~\ref{eq:unbeamed} to the {\it Fermi} Doppler
boosted LF, it is possible to determine the parameters of the
intrinsic luminosity function and of the Lorentz-factor distribution. 

	We assume that the probability density distribution for $\gamma$ is a power law
of the form
\begin{equation}
P_{\gamma}(\gamma)=C \gamma^k
\end{equation}
where C is a normalization constant and the function is valid for $\gamma_1\leq \gamma \leq \gamma_2$.
Here we assume $\gamma_1=5$ and $\gamma_2=40$ as this is the range of Lorentz 
factors observed for $\gamma$-loud  blazars \cite[e.g.][]{lahteenm03,lister09,savolainen10}. While the largest intrinsic luminosity ($\mathcal{L}_2$)
can be set free, the lowest one depends on the value of $p$ chosen: i.e.
from Eq.~\ref{eq:int}
the beaming factor defines the intrinsic luminosity corresponding to the 
apparent isotropic luminosity we observe. For p=4 and p=5, $\mathcal{L}_1$
has to be set 10$^{40}$\,erg s$^{-1}$ and 10$^{38}$\,erg s$^{-1}$ respectively.
We set $\mathcal{L}_2=10^4\mathcal{L}_1$, but this choice has hardly any impact
on the results.

\begin{figure}[h!]
\begin{centering}
	\includegraphics[scale=0.7]{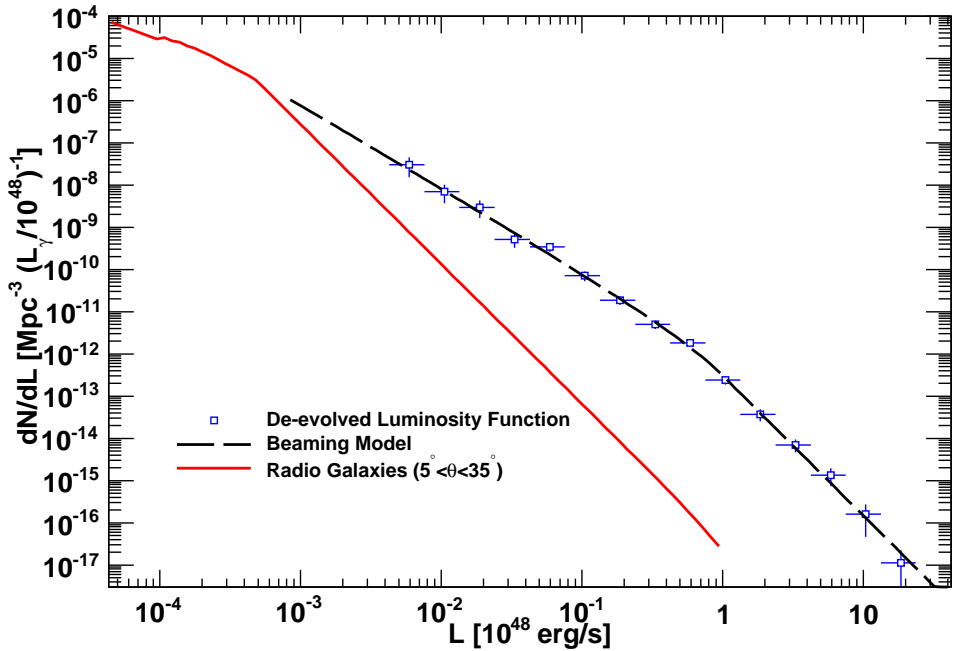} 
	\caption{{\it Fermi}'s LF de-evolved at redshift zero and the best-fit 
beaming model (for $p$=4, see text) described in $\S$~\ref{sec:beaming}. The red continuous
line shows the predicted space density of misaligned jets.
\label{fig:beaming}
}
\end{centering}
\end{figure}

The free parameters of the problem are: the normalization ($k_1$) and the slope
($B$) of the intrinsic LF and the slope $k$ of the Lorentz factor distribution.
We have fitted Eq.~\ref{eq:unbeamed} to the {\it Fermi} LF de-evolved at 
redshift zero derived in $\S$\ref{sec:local}.
Fig.~\ref{fig:beaming} shows how the best-fit beaming model reproduces the 
local LF of FSRQs measured by {\it Fermi}. From the best-fit we derive,
for the $p=4$ case, an
intrinsic LF slope of $B=3.04\pm0.08$ and an index of the 
 Lorentz-factor distribution of
 $k_1=-2.03\pm0.70$. 
The fit values are summarized in Table~\ref{tab:beaming}. 
The Lorentz-factor distribution implies an average Lorentz factor of detected 
{\it Fermi} blazars of $\gamma=11.7^{+3.3}_{-2.2}$, in reasonable 
agreement with measured values
\cite[see e.g.][]{ghisellini09}.
The index of the Lorentz-factor distribution  is in agreement
with $k_1\sim$-1.5 found for the CJ-F survey \citep{lister97}. 
The parameters for the $p=$5 case are very similar to those
of the $p=$4 case, but the reduced $\chi^2$
is slightly worse (see Table~\ref{tab:beaming}).
Nevertheless, the predictions of the two models are in agreement and
we find again that the average bulk Lorentz factor
is $\gamma=10.2^{+4.8}_{-2.4}$. As noted already the extreme Doppler
boosting ($\delta^5$) requires the intrinsic luminosities to be small: i.e. 
$10^{38}$\,erg s$^{-1}\leq\mathscr{L}\leq$10$^{42}$\,erg s$^{-1}$.

\begin{deluxetable}{lc|c}
\tablewidth{0pt}
\tablecaption{Parameters of the beaming models described in the text. 
Parameters without
an error estimate were kept fixed during the fitting stage.
\label{tab:beaming}}
\tablehead{
\colhead{Paremeter} & \colhead{Value} & \colhead{Value}}
\startdata
$k$              &  -2.03$\pm0.70$  & -2.43$\pm0.11$ \\
$k_1$            &   5.1$\pm0.5$\tablenotemark{a} & 5.0$\pm0.5$\tablenotemark{b}\\
$B$              &   3.04$\pm0.08 $ & 3.00$\pm0.08$\\ 
$\gamma_1$       &   5  & 5 \\
$\gamma_2$       &   40 & 40 \\
$\mathscr{L}_1$  & 10$^{40}$ & 10$^{38}$\\
$\mathscr{L}_2$  & 10$^{44}$ & 10$^{42}$\\
$p$              &   4 & 5\\
$\chi^2/dof$  & 1.3& 1.5\\

\enddata

\tablenotetext{a}{In units of 10$^{-23}$.}
\tablenotetext{b}{In units of 10$^{-26}$.}
\end{deluxetable}

\begin{figure}[h!]
\begin{centering}
	\includegraphics[scale=0.7]{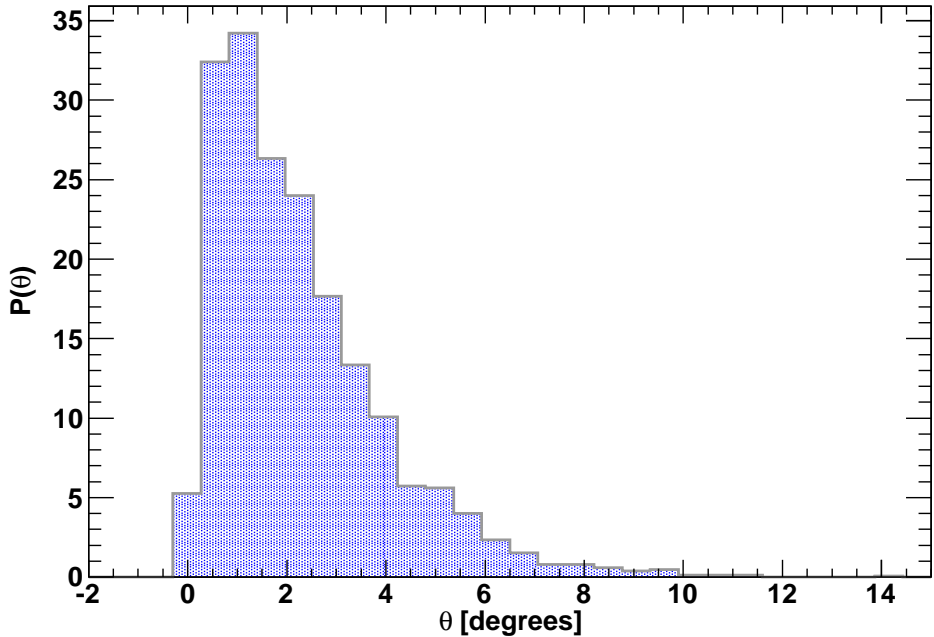} 
	\caption{Distribution of viewing angles with respect to
the jet axis for  {\it Fermi} FSRQs.
\label{fig:los}
}
\end{centering}
\end{figure}

From the ratio of the integrals of the beamed and intrinsic LF we derive
that the {\it Fermi} FSRQs represent only 0.1\,\% of the parent population.
The average space density of  LAT FSRQs (derived from the LF $\S$~\ref{sec:ldde})
is 1.4\,Gpc$^{-3}$, implying that the average space density of the
parent population is $\sim$1500\,Gpc$^{-3}$.  Our model also allows us to 
determine the distribution of jet angles with respect to our line of 
sight. This is found to peak at $\sim$1.0\,degrees (Fig.~\ref{fig:los}).
While FSRQs can still be detected at large (i.e. $\geq$10\,degrees)
off-axis angles for reasonably low $\gamma$ factors ($\sim$5-7),
most  FSRQs detected by {\it Fermi} are seen at angles less 
than 5\,degrees from the jet axis. Owing to their larger space
density (see Fig.~\ref{fig:beaming}) misaligned jets produce
a non-negligible diffuse emission. From our model it is found
that the ratio between the diffuse emission contribution of misaligned jets
and that of blazars (at redshift zero) is $\sim$30\,\%. 
This has obvious consequences
for the generation of the IGRB. In fact nearly all of the flux produced by 
radio galaxies is unresolved, with only 2 steep-spectrum radio quasars,
and 2 FR II and 7 FR I radio galaxies detected with the {\it Fermi} LAT
\citep{grandi10}. All the results reported above apply to both the $p=4$ and $p=5$ models.

%
%
\section{Discussion and Conclusions}

	In this paper we examine the properties of $\gamma$-ray selected 
FSRQs using data from the {\it Fermi}-LAT instrument. Our work relies on 
a nearly complete, flux-limited sample of 186 FSRQs detected by 
{\it Fermi} at high significance and large Galactic latitude during the first 
year of operations. This analysis explores several of the properties of 
FSRQs; here we discuss and summarize our findings.

\subsection{Beamed Luminosity Function}
The redshift-zero LF of {\it Fermi} FSRQs is well described by a double power-law model, typical for the LF of AGN (both of the radio-quiet and radio-loud).
At mid-to-high luminosities there is good agreement between the {\it Fermi} LF 
and that determined using EGRET data \citep[e.g.][]{narumoto06,inoue10b}. At 
luminosities $\leq10^{46}$\,erg s$^{-1}$ the FSRQ LF appears to be slightly 
flatter than in previous studies.

The space density of LAT-detected 
FSRQs increases dramatically with redshift, growing
by 50--100$\times$ by z=1.5. Describing the evolution of the LF as simple 
luminosity evolution (PLE model), there are strong indications that the 
evolution must cut off for z$\geq1.6$. After this redshift, the space density 
of blazars starts to decrease quickly.

A simple PLE does not fully explain the {\it Fermi} data. In particular
the source count distribution is not well modeled. Since there is evidence
that low- and high-luminosity sources have different redshift peaks, we
consider a more sophisticated model where the evolution is primarily in density,
but objects with different luminosity are allowed to have different redshift
peaks. This so-called LDDE model explains well the evolutionary behavior 
of (radio-quiet) AGN selected in the X-ray band \citep[][]{ueda03,hasinger05}
and was also suggested by \cite{narumoto06} to describe the LF of EGRET blazars.
The LDDE model provides a good description of the LF of the {\it Fermi} FSRQs.  
We find that  the  predictions
reported in the literature \cite[e.g.][]{narumoto06,inoue10b,stecker11} 
are not in agreement with the LF of {\it Fermi} FSRQs at redshift unity.
This is due to the fact that the {\it Fermi} FSRQs  is found 
to evolve more quickly than the LFs of  X-ray-selected AGN or
 radio-selected FSRQs. Indeed, the space density of {\it Fermi} FSRQs
increases by a factor $\sim$150 between redshift 0 and 1 while the 
density increase is at most a factor $\sim$60 for the models discussed above.

The LDDE model implies that sources with a luminosity of 10$^{46}$\,erg s$^{-1}$,
10$^{47}$\,erg s$^{-1}$, and 10$^{48}$\,erg s$^{-1}$ reach their maximum
space density at a redshift of $\sim$0.6, $\sim$0.9, and $\sim$1.5 respectively.
It is clear, then, that the most luminous objects, while lower in numbers,
appear before the bulk of the (low-luminosity) population. 
This is the first time that this is seen in $\gamma$-rays.
This ``anti-hierarchical''  behavior, where the largest structures come first
is common to all classes of  AGN \cite[see e.g. ][ and references
therein]{cowie99,hasinger05}, and is often referred to as
``cosmological downsizing''. 
The disappearance of quasar-like objects 
at late times might indicate that accretion efficiency evolves
as a function of cosmic time \cite[e.g.][]{merloni04}. \cite{dimatteo05} 
\citep[but see also ][]{sanders88}
propose that the merging of two massive galaxies leads to, in addition to 
strong star formation activity, a burst of inflow feeding gas to the SMBH
and initiates a 'quasar-like' phase.  Eventually the energy released by the 
AGN in the form of powerful winds expels the gas, quenches star formation and starves the
AGN. This picture, coupled with the fact that major mergers become increasingly rare
at low redshift \cite[e.g.][]{fakhouri10,kulkarni11} may explain why
quasars are rare in the local Universe.

\begin{figure}[h!]
\begin{centering}
	\includegraphics[scale=0.7]{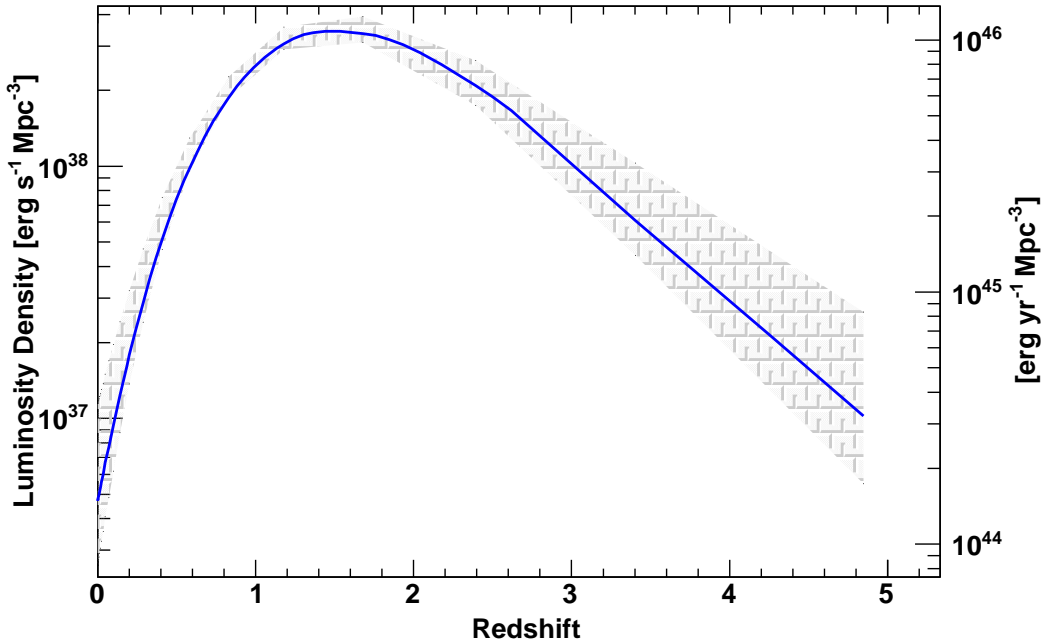} 
	\caption{Luminosity density as a function of redshift produced by
the {\it Fermi} FSRQs.
 The gray band represents the 1\,$\sigma$ statistical uncertainty
around the best-fit LF model.
	\label{fig:lumin_density}
}
\end{centering}
\end{figure}

Fig.~\ref{fig:lumin_density} shows the energy density injected in the
Universe (e.g. the luminosity density) by FSRQs as function of redshift.
This shows a broad peak between a redshift of 1 and 2. A similar behavior
is shown by the cosmic star-formation history \cite[e.g.][]{hopkins06}
which peaks around redshift 1-2. This represents a strong link
between the host and the nucleus.  A noteworthy fact is the mild
evidence for a fast decline in the space density of FSRQs after the redshift peak 
(see parameter $p_2$ in Table~\ref{tab:ldde}). The decline seems to be as dramatic 
as the increase in space density leading up to the redshift peak. For comparison, 
X-ray selected samples of AGN show a much milder decline ($p_2\approx-1.5$)
after the redshift peak \citep[e.g.][]{ueda03,hasinger05,aird10}.
However, recently \cite{silverman08} \citep[but see also][]{schmidt95} reported
evidence for a similarly dramatic decrease in the space density of AGN.

One factor contributing to this phenomenon is the difficulty for {\it Fermi} 
to detect soft sources \citep{pop_pap}. At redshift $\geq$3 the SED peak should 
move well below the current LAT energy band, making it difficult to probe a
population of extremely soft sources. 
Increasing the effective area at or below 100\,MeV may help
uncover such a population.
Because the rising part of the IC peak is in the hard ($\geq$10\,keV) X-ray band,
high-redshift objects are more easily selected in this band 
(see e.g. the {\it Swift}/BAT results in \cite{ajello09b}). In this case
another strategy would be to build a bolometric luminosity function
that uses both the $\gamma$-ray and the X-ray selected samples.

\begin{figure}[h!]
\begin{centering}
	\includegraphics[scale=0.7]{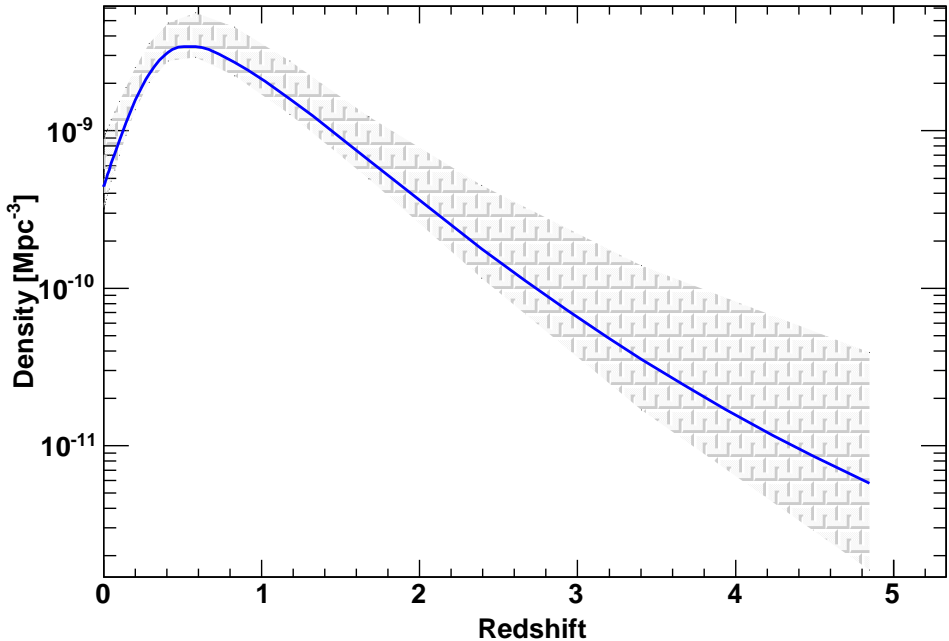} 
	\caption{Number density of LAT-detected FSRQs as a function of redshift.
 The gray band represents the 1\,$\sigma$ statistical uncertainty
around the best-fit LF model.
	\label{fig:density}
}
\end{centering}
\end{figure}

%
%
\subsection{The Intrinsic Luminosity Function}

Doppler boosting allows {\it Fermi} to detect many blazars when
their jet emission is within a few degrees from the line of sight.
As shown first by \cite{urry84}, Doppler boosting is known to alter
the shape of the luminosity function. In this paper, we adopted
a formalism that allowed us to recover the intrinsic de-beamed LF
and to determine the distribution of bulk Lorentz factors
for the {\it Fermi} FSRQs.

The intrinsic LF is compatible with a single steep power law with
an index of 3.04$\pm0.08$ in the range of intrinsic luminosities
$10^{40}$\,erg s$^{-1}$$\leq \mathscr{L}\leq 10^{44}$\,erg s$^{-1}$.
The break seen in the {\it beamed} LF at redshift zero is thus
produced by Doppler boosting.
The data cannot be explain by a single, averaged, Lorentz factor, but
require a distribution of Lorentz factors.
This distribution is found to be  compatible with a power law with an 
index of -2.03$\pm0.70$ in the 5$\leq$$\gamma$$\leq$40 range.
This yields the result that the average FSRQ bulk Lorentz factor is 
$\Gamma=11.7^{+3.3}_{-2.2}$,
in good agreement with several studies \citep{ghisellini09}.
Our model is able to predict the distribution of viewing angles
with respect to the jet axes of {\it Fermi} FSRQs.
It is found, see Fig.~\ref{fig:los}, 
that on average FSRQs are seen within an average
angle of $\sim$2.3$^{\circ}$ from the jet and that most are seen
within 5$^{\circ}$-6$^{\circ}$. A few 
{\it Fermi} FSRQ detections may be up to 15$^{\circ}$ off-axis
(if these have low Doppler factors). {\it Fermi}-detected FSRQs 
represent only $\sim$0.1\,\% of the parent population for randomly 
pointed jets.

Monitoring observations with the Very Long Baseline Array (VLBA)
established that LAT detected blazars have, on average,
significantly faster apparent jet speeds than non-LAT detected blazars
\citep{lister09b,savolainen10}. Their distribution of Lorentz factors
is in good agreement with the results of our analysis.
Moreover, they report the distribution of viewing angles with respect
to the jet axis for FSRQs detected by LAT. From their study 
the average viewing angle is 
$2.9^{\circ}\pm0.3^{\circ}$ and all the FSRQs in their sample have
$\theta\leq5^{\circ}$. There is thus substantial agreement within the VLBA
monitoring observations and the results of our beaming model applied
to $\gamma$-ray only data.

The space density of FR-II radio galaxies (i.e. the
putative parent population of FSRQs) is reported to be  
$\sim$1580\,Gpc$^{-3}$ (at 15\,GHz) and $\sim$2200\,Gpc$^{-3}$ (at 1.4\,GHz)
by \cite{cara08} and \cite{gendre10} respectively.
From our study we derive a space density of FSRQ parents of 
$\sim$1500\,Gpc$^{-3}$ in substantial agreement with the numbers above.

Future work may test whether the intrinsic properties of blazars 
(i.e. Lorentz factor, luminosity etc.) evolve with redshift. This 
will likely require larger  complete samples and improved modeling
of selections effects.

%
%
\subsection{Spectral Energy Distribution}
Blazars SEDs are characterized by the typical ``two hump'' spectrum
where the low-energy peak is produced by electrons radiating via
synchrotron and the high energy peak is produced via IC scattering 
off the same synchrotron photons \citep[synchrotron-self Compton scenario; ][]{maraschi92}
and/or external seed photons \citep[external Compton scenario;][]{dermer93}.

In this work we have combined quasi-simultaneous {\it Swift}/BAT and 
{\it Fermi}/LAT data to investigate the empirical properties of the IC
component of the SEDs of the FSRQs detected by {\it Fermi}.
All the SED show apparent curvature and have a peak somewhere
in the 10\,MeV--1\,GeV band. There is no correlation between
the IC peak luminosity and energy for the sample of FSRQs detected by
{\it Fermi}. The existence of such correlation has been
claimed in the past for the luminosity and the energy of the
synchrotron peak  \citep{ghisellini98,fossati99} for a sample of
blazars (i.e. FSRQs and BL Lacs). 
Thus it might be that this
correlation (if real) exists only when the two families
of blazars are joined together and that any correlation for the
FSRQs class is washed away by the presence of the additional
EC component. Also the lack of correlation of the IC peak luminosity and
frequency reveals that FSRQs are, unlike BL Lacs, part of a population with homogenous properties.

We built average redshift-corrected SEDs in four different luminosity bins.
The average SEDs are surprisingly similar as a function of luminosity
(and redshift) as Fig.~\ref{fig:averagesed} testifies.
Approximating the SED with a power law with an index 2.4, while
not producing the correct shape, allows the reader to compute
a k-correction useful up to redshift $\approx$2. Beyond that this
approximation is not valid.

%
\subsection{The Contribution to the Diffuse Background}

This work has important consequences for our understanding of the origins
of the
diffuse background. As pointed by several authors \citep[e.g.][]{inoue09} and
determined in this work, the spectrum of the diffuse emission	
arising from FSRQs shows curvature, due to the curved SEDs of these objects. 
We couple our model SED to our LF to predict the FSRQ contribution to the 
10\,keV to 100\,GeV diffuse background.  FSRQs produce $\sim$10\,\% 
of the cosmic diffuse emission in the 1\,MeV-10\,GeV band.

Because of its good sensitivity {\it Fermi} has already resolved
as much as 50\,\% of the total flux from FSRQs in the 100\,MeV--100\,GeV 
band. Our analysis indicates that the contribution of {\it unresolved}
FSRQs to the IGRB \citep{lat_edb} is 9.66$^{+1.67}_{-1.09}\times10^{-7}$\,ph cm$^{-2}$ s$^{-1}$ sr$^{-1}$
and thus only  9.3$^{+1.6}_{-1.0}$\,\% ($\pm$3\,\% systematic uncertainty)
of the intensity of the IGRB.
This analysis is in good agreement with the results reported by \cite{pop_pap}
except above 10\,GeV where the use of a simple power law for the spectra
of FSRQs was inadequate.

Our results appear in reasonably good agreement with those of \cite{inoue10b} and
of \cite{inoue11}, both in terms of spectral shape of the diffuse emission
arising from FSRQs and its intensity. In their work, these authors rely
on the sample of FSRQs and BL Lacs detected by EGRET. It is thus not
surprising that their estimates of the contribution to 
the IGRB are slightly larger than ours. Finally, our estimate reported above
is in good agreement with the results of \cite{dermer07} that predicted
that FSRQs would produce $\approx$10--15\,\% of the $\gamma$-ray background.

LAT-detected FSRQs represent only $\sim0.2$\,\% of the parent population 
(see $\S$~\ref{sec:beaming}) and thus it is reasonable to expect
that misaligned jets, although less luminous, but more numerous
give a non-negligible contribution to the diffuse background.
Our beaming model allowed us to explore this scenario.
It is found that misaligned relativistic jets contribute $\sim$30\,\%
of the diffuse flux from the FSRQs class at redshift zero. 
If the Lorentz factor distribution does not change with redshift then
the contribution of {\it unresolved} FSRQs and their misaligned
siblings might be around $\sim$2.0$\times10^{-6}$\,ph cm$^{-2}$ s$^{-1}$ 
and thus $\sim$20\,\% of the IGRB.
Recently, \cite{inoue11} predicted that radio galaxies of both
the FR-I and FR-II type might be able to account for $\sim$25\,\%
of the intensity of the IGRB. In our work we found that FR-II alone
could in principle (see above caveat) produce $\sim$10\,\% of the IGRB. It can be envisaged
that once also the contribution to the IGRB of BL Lacs and their
parents will be established, the total $\gamma$-ray emission
from relativistic jets might account for some $\sim$40--50\,\% of the intensity
of the IGRB.

%
%

\clearpage
\appendix
\section{Systematic Uncertainties}
\label{app:sys}

The sources of systematic uncertainties in this analysis are:
incompleteness (i.e. missing redshifts), detection efficiency,
blazar variability, and EBL. A detailed discussion of some of these problems
was already given in \cite{pop_pap}. Incompleteness in our sample is very small 
and introduces no appreciable systematic uncertainty as shown in $\S$~\ref{sec:sys}.

\subsection{Detection Efficiency}
\begin{figure}[h!]
\begin{centering}
	\includegraphics[scale=0.7]{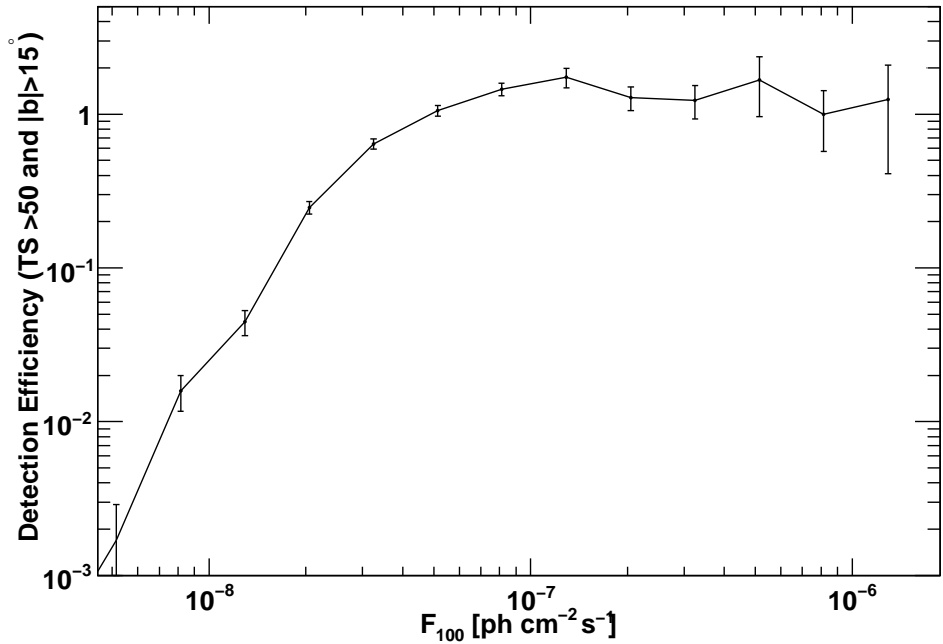} 
	\caption{Detection efficiency as a function of flux for a population
of sources with curved spectra similar to those of FSRQs determined
in $\S$~\ref{sec:sed}.
	\label{fig:det}
}
\end{centering}
\end{figure}

The detection efficiency used to determine the sky area surveyed
by {\it Fermi} at any given flux is very important in this analysis
\citep[see][for a detailed discussion]{pop_pap}.
The detection efficiency used in this work was derived in \cite{pop_pap}
under the assumption that the blazars spectra can be approximated
by a power law. While this might be true over a small
energy band, it becomes a problematic assumption over the full 100\,MeV--100\,GeV
band covered by the LAT. In this Section we estimate directly the systematic uncertainties
connected to this hypothesis. We performed 3 end-to-end Monte Carlo 
simulations of the {\it Fermi} sky \citep[see ][for details]{pop_pap},
assigning randomly a curved spectrum to each source.
These spectra are extracted from a library created using the
$\sim$100 observed spectra derived in $\S$~\ref{sec:sed} varying
the parameters of the measured spectra within their errors.
The simulations were then analyzed to derive the detection efficiency
reported in Fig.~\ref{fig:det}.  In particular in order to detect
a source a maximum likelihood fit with a power law spectrum is performed.
This is done in order to reproduce the inherent systematic
uncertainty of fitting the curved spectrum of a source with a power law
\citep[][]{cat1}.

Fig.~\ref{fig:det} shows the detection efficiency for a sample
like that used in this analysis. Two aspects are noteworthy.
First the efficiency at F$_{100}=10^{-8}$\,ph cm$^{-2}$ s$^{-1}$
(i.e. the lowest flux of this analysis by construction) is $\sim0.02$
with a typical uncertainty of $\pm5\times10^{-3}$ 
dictated by the small statistic in our simulations at the lowest fluxes.
Second, at fluxes around F$_{100}\approx 10^{-7}$\,ph cm$^{-2}$ s$^{-1}$
the detection efficiency becomes larger than 1.0.
This effect is due to the fact that fitting a curved spectrum source
with a power law yields to an overestimate of the source flux
by a factor $\sim10$\,\% \citep[see also Fig.~8 in ][]{pop_pap}.
Since Fig.~\ref{fig:det} is built as the ratio (in a given bin)
of the number of sources detected with a given flux to the number
of simulated sources with that flux, the effect mentioned above
leads to  a detection efficiency $>$1.0.

In order to test the level of systematic uncertainty we derived the LF
using the detection efficiency reported in Fig.~\ref{fig:det}. Given
the ``small'' number of sources detected in the 3 simulations, it was not
possible to derive a two-dimensional detection efficiency as a function
of flux and spectral index \citep[like that used in 
$\S$~\ref{sec:results} and derived for power law sources in ][]{pop_pap}.
For this reason the parameters of the distribution of photon indices 
of the FSRQ class cannot be derived from the analysis of the LF.
As it is apparent from Table~\ref{tab:ldde} most parameters
of the LF derived in this section and those derived in $\S$~\ref{sec:ldde}
are compatible within their statistical errors.
The only parameter for which the difference is slightly larger than the 
statistical errors is $\alpha$. The parameter $\alpha$
governs the trend of the redshift peak with luminosity and while
its statistical error is in both case 0.03, the systematic error appears
to be 0.05. This has very little impact on the analysis and the
results of the previous sections are fully confirmed and robust
against variations of the detection efficiency curve.
As a further proof, the points of the de-evolved LF in 
Fig.~\ref{fig:local} and Fig.~\ref{fig:z=1} were computed
using the detection efficiency of Fig.~\ref{fig:det} while the
shaded error region was computed using the model LF
derived in $\S$~\ref{sec:ldde} that uses the detection efficiency
for power law source.

We performed an additional test, by shifting the detection
efficiency curve in Fig.~\ref{fig:det} to fluxes 10\,\% brighter 
than measured. The rightward shift is most dramatic as it increases
the magnitude of the correction at faint fluxes. The shift is performed in
order to account for uncertainties in the determination of
the detection efficiency. The parameters of the LF are all
consistent within statistical uncertainty with those found in
this and the previous sections and reported in Table~\ref{tab:ldde}.
The index of the low-luminosity slope of the LF becomes slightly steeper
(i.e. $\gamma_1$=0.47$\pm0.18$), and this yields a slightly larger
contribution to the IGRB from FSRQs. We thus consider the typical 
systematic uncertainty connected to the estimate of the contribution to the IGRB
to be $\sim$3\,\% of the IGRB 100\,MeV--100\,GeV intensity.

\subsection{Variability}
It is well known that blazars are inherently variable objects with variability
in flux of up to a factor 10 or more. Throughout  this work
only average quantities (i.e. mean flux, mean luminosity and
 mean photon index) are used.

It is not straightforward to determine how blazar variability affects
the analysis presented here. While the variability patterns and amplitudes
of blazars as a class are still not known
 both \cite{lbas_variability} and \cite{2LAC}
presented a detailed analysis of the variability of the brightest
{\it Fermi} blazars. They report that the variability amplitude
of the FSRQ class is generally larger than that of the BL Lac population.
However, most sources (either bright or faint ones)
exceed their average flux for less than 5--20\,\% 
of the monitored time (i.e. respectively 11\,months or 2\,years).
This drastically reduces the possibility that FSRQs (or blazars more in general)
are detected because of a few bright flaring episodes.
The effect of high-amplitude variability connected with a rising density of 
sources at smaller flux might contaminate samples, as the one used here,
with objects that formally should not be included. However, because of what
reported above and the flatness of the FSRQs source count distribution \citep[see ][ and Fig.~\ref{fig:ldde}]{pop_pap,2LAC} this effect is very likely marginal.

Another, smaller, problem is connected to the dependence of the effective
area on  the direction of the incoming photon
and the LAT detector frame\footnote{See e.g. http://www.slac.stanford.edu/exp/glast/groups/canda/lat\_Performance.htm and \cite{atwood09}.}.
Short intense flares detected during favorable conditions (i.e. on axis
and at azimuthal angles of $\sim$0, $\sim$90,  $\sim$180, or $\sim$270 degrees)
might lead to a higher TS, increasing the likelihood of  source detection.
However, because of {\it Fermi}'s continuous scanning of the sky and because most flares are observed to last 10\,days or longer \citep{lbas_variability}, the effect above has negligible influence on the analysis presented here.

Finally, we believe that variability does not hamper the results of this analysis.
Using even longer integration times (e.g. 2, or 3\,years) will be the most
efficient way to confirm the results of this analysis and dilute the effect
of blazar variability.

\subsection{Extragalactic Background Light}
Uncertainty in the level of the EBL, in particular at medium to high redshift,
might in principle affect our analysis. Energetic $\gamma$-rays from FSRQs at
high redshifts might be absorbed by the EBL and if this effect is not taken
into account the source-frame luminosity would be underestimated. This would lead
to wrong estimates of the  space densities of FSRQs. However we believe this 
uncertainty is negligible. 

The uncertainty in the level of the EBL would impact the estimate of the
k-correction which allows us to determine the source-frame luminosities.
As shown in Fig.~\ref{fig:kcorr}, neglecting the EBL at once and adopting
a simple power-law spectrum with a photon index of 2.4 (instead of the
average SED determined in $\S$~\ref{sec:sed}) introduces an uncertainty
of $\leq$4\,\% on the value of the k-correction at z=3. Since all
the {\it Fermi} FSRQs are detected within this redshift, this uncertainty
produces no appreciable impact on the determination of the luminosity function.

\clearpage
\acknowledgments
MA acknowledges Y. Inoue and  T. Venters for providing their data in 
electronic form and for interesting discussions about the origin of the IGRB.
MA acknowledges support from  NASA grant NNH09ZDA001N for the study
of the origin of the IGRB. RWR acknowledges NASA grant NNX08AW30G and
extensive consultation with the OVRO Fermi group.

The \textit{Fermi} LAT Collaboration acknowledges generous ongoing support
from a number of agencies and institutes that have supported both the
development and the operation of the LAT as well as scientific data analysis.
These include the National Aeronautics and Space Administration and the
Department of Energy in the United States, the Commissariat \`a l'Energie Atomique
and the Centre National de la Recherche Scientifique / Institut National de Physique Nucl\'eaire et de Physique des Particules in France, the Agenzia 
Spaziale Italiana and the Istituto Nazionale di Fisica Nucleare in Italy, 
the Ministry of Education, Culture, Sports, Science and Technology (MEXT), 
High Energy Accelerator Research Organization (KEK) and Japan Aerospace 
Exploration Agency (JAXA) in Japan, and the K.~A.~Wallenberg Foundation, 
the Swedish Research Council and the Swedish National Space Board in Sweden.
Additional support for science analysis during the operations phase 
is gratefully acknowledged from the Istituto Nazionale di Astrofisica in 
Italy and the Centre National d'\'Etudes Spatiales in France.

{\it Facilities:} \facility{Fermi/LAT}, \facility{Swift/BAT}

\bibliographystyle{apj}
\bibliography{/Users/majello/Work/Papers/BiblioLib/biblio.bib}

\end{document}